\documentclass[12pt]{article}
\usepackage[table]{xcolor}
\usepackage{graphicx}%
\usepackage{multirow}%
\usepackage{amsmath,amssymb,amsfonts}%
\numberwithin{equation}{section}%
\usepackage{amsthm}%
\usepackage{mathrsfs}%
\usepackage{cite}
\usepackage[title]{appendix}
\usepackage{xcolor}%
\usepackage[colorlinks=true, linkcolor=blue, citecolor=blue]{hyperref}
\usepackage{textcomp}%
\usepackage{manyfoot}%
\usepackage{threeparttable}
\usepackage{booktabs}%
\usepackage{footnote}
\usepackage{algorithm}%
\usepackage{algorithmicx}%
\usepackage{algpseudocode}%
\usepackage{listings}%
\usepackage{titlesec}
\usepackage[utf8]{inputenc}
\usepackage{slashed}
\usepackage{hyperref}
\usepackage{url}
\usepackage{doi}
\hypersetup{colorlinks=true,linkcolor=blue,urlcolor=blue}
\usepackage[framemethod=TikZ]{mdframed}
\mdfdefinestyle{MyFrame}{%
    linecolor=blue,
    outerlinewidth=2pt,
    roundcorner=20pt,
    innertopmargin=\baselineskip,
    innerbottommargin=\baselineskip,
    innerrightmargin=20pt,
    innerleftmargin=20pt,
    backgroundcolor=white!50!white }
\begin{document}
\title{Anti de Sitter gauge theory for gravity}
\author{Theo Verwimp\footnote{PhD graduate from U.I.A., Universiteit Antwerpen Belgium. On retirement from ENGIE Laborelec, Belgium}.\\ {\small ORCID} \url{https://orcid.org/0009-0001-3086-0445}}
\maketitle
First a review is given of Riemann-Cartan space-time and Einstein-Cartan gravity. This gives us the necessary tools to handle the SO(2,3) Yang-Mills gauge theory for gravity. Field equations are obtained from a Yang-Mills gauge field Lagrangian. From these field equations and the Bianchi identities the conservation laws for anti-de Sitter theory are derived. Possible solutions of the field equations are discussed. The only solution found for a RW-geometry with a perfect fluid matter content is the early universe. There are no Schwarzschild solutions of the torsion free vacuum field equations if curvature squared terms can  not be neglected.
\tableofcontents
\section{Introduction}\label{sec.1}
The fact that gravity can be described as an anti-de Sitter gauge field theory has been known for some time. Gravity was first formulated as a de Sitter or anti-de Sitter gauge theory in papers by Townsend, MacDowell and Mansouri, Smrz, and Stelle and West \cite{bib1,bib2,bib3,bib4}. Later, also Zardecki used the notion of the connection of Cartan to describe gravity in gauge formalism \cite{bib5}.  Gotzes and Hirshfeld used a Clifford algebra valued Cartan connection to give a geometical formulation of anti-de Sitter gravity \cite{bib6}. Also Ashtekar's formalism \cite{bib7}, was applied to the gauge theory of the de Sitter group.
     
     In order to reproduce the structure of an Einstein-Cartan theory, the $SO(2,3)$ gauge symmetry must be spontaneously broken to the Lorentz-symmetry. The symmetry breaking mechanism with the Higgs scalar field was discussed extensively by Stelle and West \cite{bib4}. The geometrical concepts involved in the process of symmetry breaking in electroweak theory were compared with the corresponding geometrical concepts in anti-de Sitter gauge theory in \cite{bib8}. How the anti-de Sitter gauge theory can arise in unification of the fundamental interactions, see Refs \cite{bib9,bib10}.

In this work, we consider the anti-de Sitter gauge theory for gravity as a pure Yang-Mills theory. In section \ref{sec.2} we present the basic information on the structure of a Riemann-Cartan spacetime (see also Refs \cite{bib11,bib12,bib13,bib14,bib15,bib16}). This is a Riemannian manifold with a solder form $\theta^{a}$ and a metric compatible spin connection $\omega^{ab}$ as fundamental geometrical variables. In terms of these variables a complete and intrinsic description is given of the gravitational field on the spacetime manifold.      
    
In an anti-de Sitter gauge theory with symmetry breaking to $SO(1,3)$, gravitational interaction must be formulated in terms of the local Lorentz group. Therefore in section \ref{sec.3} we consider local Lorentz transformations for matter fields. The Noether identity for this Lorentz symmetry is derived for a massive fermion field. Expressions for the energy-momentum and spin angular-momentum 3-forms are obtained using the expression for a general variation of the matter Lagrangian. In section \ref{sec.4} we consider the gauge field Lagrangian for a local Lorentz gauge theory and derive the field equations for Einstein-Cartan gravity. Then using the first Bianchi-identity, the Noether identity for the local Lorentz symmetry derived in section \ref{sec.3}, is obtained from the field equations. Using both Bianchi-identities and the field equations a conservation law for the energy-momentum is derived.
    
     Starting from the Yang-Mills action for the $SO(2,3)$ symmetry defined on the anti-de Sitter frame bundle, we obtain in section \ref{sec.5} by re-expressing the gauge field Lagrangian in terms of quantities defined on the Lorentz-frame subbundle, the expression for the gravitational Lagrangian in anti-de Sitter gauge theory. Subsequently, the gravitational field equations are derived and the conservation laws for anti-de Sitter gravity. These are the Noether identity for the local Lorentz symmetry from section \ref{sec.3} and \ref{sec.4} and the conservation law for the energy-momentum which is of the same form as for Einstein-Cartan gravity. 
     
     Solutions of the field equations depend on the scale of the de Sitter length $l$. Different theories are obtained depending on the scale length. Explicit solutions in case of a Schwarzschild and RW-geometry are given in the appendices.
\section{Riemann-Cartan Spacetime}\label{sec.2}
\subsection{The Vierbein}
Let $(M,g)$ be a four-dimensional, paracompact, Hausdorff, connected $C^{\infty}$-manifold, with a non-singular metric $g$ of signature (-,+,...,+).
      
      If $\lbrace X_{a}(x)\rbrace_{a=0,1,2,3}$ is an orthonormal frame, and $\lbrace\partial_{\alpha}(x)\rbrace_{\alpha=0,1,2,3}$ a coordinate basis, then the $ X_{a}$  transform under $ SO(1,3)_{local}$ into an orthonormal basis, while the $ \partial_{\alpha}$ transform under $ GL(4,\mathbb{R})_{local}$.
      
      Let $e^{\alpha}_{a}$ be the matrix which transforms the coordinate basis ${\partial_{\alpha}}$ into the orthonormal basis ${X_{a}}$:
\begin{equation}
X_{a}=e^{\alpha}_{a}\partial_{\alpha}\label{eq2.1}
\end{equation}
and denote by $e^{a}_{\alpha}$ the inverse transformation
\begin{equation}
\partial_{\alpha}=e^{a}_{\alpha}X_{a}\label{eq2.2}
\end{equation}
With,
\begin{equation}
g=g^{\alpha\beta}{\partial_{\alpha}}\otimes{\partial_{\beta}}={\eta^{ab}X_{a}}\otimes X_{b},\quad\eta=diag(-1,+1,...,+1)\label{eq2.3}
\end{equation}
it follows that
\begin{equation}
g^{\alpha\beta}=\eta^{ab}e^{\alpha}_{a}e^{\beta}_{b},\qquad\eta^{ab}=g^{\alpha\beta}e^{a}_{\alpha}e^{b}_{\beta}\label{eq2.4}
\end{equation}
   Since in each point $x\in M$, the $e^{\alpha}_{a}$ are the components of a basis of orthonormal tangent vectors, these components are also called \emph{vierbein}. The Lorentz index $a$ of the vierbein is raised and lowered with the Minkowski-metric $\eta_{ab}$, while the spacetime index $\alpha $ is moved with $g_{\alpha\beta}$ ($a$ is a SO(1,3) tensor index and $\alpha$ a $GL(4,\mathbb{R})$ tensor index). Therefore, we have from (\ref{eq2.4})
   \begin{equation}
   e^{\alpha}_{a}e^{a}_{\beta}=\delta^{\alpha}_{\beta},\qquad e^{a}_{\alpha}e^{\alpha}_{b}=\delta^{a}_{b} \label{eq2.5}
   \end{equation}
         
         If $\theta^{a}(x)$ are the dual one-forms corresponding with $ X_{a}(x)$, this means, $\theta^{a}(X_{b})=\delta^{a}_{b}$, then it follows that
     \begin{equation}
         \theta^{a}=e^{a}_{\alpha}dx^{\alpha}\label{eq2.6} 
     \end{equation}
     \begin{equation}
     dx^{\alpha}=e^{\alpha}_{a}\theta^{a}\label{eq2.7}
     \end{equation}
     Also we have
     \begin{equation}
     g=g_{\alpha\beta}dx^{\alpha}\otimes dx^{\beta}=\eta_{ab}\theta^{a} \otimes \theta^{b}\label{eq2.8}
     \end{equation}
  with
  \begin{equation}
  g_{\alpha\beta}=\eta_{ab}e^{a}_{\alpha}e^{b}_{\beta},\qquad \eta_{ab}=g_{\alpha\beta}e^{\alpha}_{a}e^{\beta}_{b}\label{eq2.9}
  \end{equation}
      
      The vierbein and its inverse $e^{a}_{\alpha}(x)$, are determined only upon a local  Lorentz transformation $\Lambda(x)$
      \begin{equation}
        \Lambda:e^{a}_{\alpha}\rightarrow e'^{a}_{\alpha}=\Lambda^{a}_{b}(x)e^{b}_{\alpha}\label{eq2.10}
        \end{equation} 
    The corresponding co-frame $\theta^{a}$ transforms as
    \begin{equation}
     \Lambda:\theta^{a}\rightarrow \theta'^{a}=\Lambda^{a}_{b}(x)\theta^{b}\label{eq2.11}
     \end{equation} 
while the orthonormal basis transforms as
\begin{equation}
  \Lambda:X_{a}\rightarrow X'_{a}=X_{b}(\Lambda^{-1}(x))^{b}_{a}\label{eq2.12}
  \end{equation}
such that $\theta^{a}(X_{b})=\theta'^{a}(X'_{b})=\delta^{a}_{b}$.
 
  Analogous to (\ref{eq2.1}), (\ref{eq2.2}), (\ref{eq2.4}), (\ref{eq2.6}), (\ref{eq2.7}) and (\ref{eq2.9}), tensor indices with respect to the holonomic basis (anholonomic basis), can be transformed into tensor indices with respect to the anholonomic basis (holonomic basis), by multiplying with the matrices $e^{\alpha}_{a}(e^{a}_{\alpha})$. For example, let $V(x)$ be a vector field, then with (\ref{eq2.1})
  \begin{equation}
  V=V^{\alpha}\partial_{\alpha}=V^{a}X_{a}=V^{a}(e^{\alpha}_{a}\partial_{\alpha})\label{eq2.13}
  \end{equation}
such that
\begin{equation}
V^{\alpha}=e^{\alpha}_{a}V^{a}\label{eq2.14}
\end{equation}
and with (\ref{eq2.5})
\begin{equation}
V^{a}=e^{a}_{\alpha}V^{\alpha}\label{eq2.15}
\end{equation}

\subsection{The Spin Connection}

The covariant derivative of a vector field $ V(x)=V^{\alpha}(x) \partial_{\alpha}(x)$ with respect to the coordinate basis is given by
\begin{equation}
D_{\alpha}V=(D_{\alpha}V^{\beta})\partial_{\beta},\qquad D_{\alpha}V^{\beta}=\partial_{\alpha}V^{\beta}+\Gamma^{\beta}\medskip_{\alpha\gamma}V^{\gamma}\label{eq2.16}
\end{equation}
 with the $GL(4,\mathbb{R})$ connection defined as
\begin{equation}
 D_{\alpha}\partial_{\gamma}=\Gamma^{\beta}\medskip_{\alpha\gamma}\partial_{\beta}\label{eq2.17}
 \end{equation}
  The covariant derivative of a vector field $ V(x)=V^{a}(x) X_{a}(x)$ with respect to the orthonormal basis is given by
 \begin{equation}
D_{\alpha}V=(D_{\alpha}V^{a})X_{a},\qquad D_{\alpha}V^{a}=\partial_{\alpha}V^{a}+\omega^{a}\medskip_{b \alpha}V^{b}\label{eq2.18}
\end{equation} 
  with the $SO(1,3)$ connection defined as
  \begin{equation}
  D_{\alpha}X_{a}=\omega^{b}\medskip_{a\alpha}X_{b}=\omega^{b}\medskip_{a}(\partial_{\alpha})X_{b}\label{eq2.19}
  \end{equation}
  The matrix of 1-forms
  \begin{equation}
  \omega^{ab}=\omega^{ab}\medskip_{\alpha}dx^{\alpha}\label{eq2.20}
  \end{equation}
       is called the \emph{spin connection}.This connection can be coupled to spinor fields, which is not the case for the $GL(4,\mathbb{R})$ connection. As with other gauge fields, the spin connection can be coupled to a field in any required representation of the local Lorentz group.
      
      The requirement that the covariant derivative transforms homogeneously under a local Lorentz transformation, this is,
      \begin{equation}
      \Lambda:D_{\alpha}V^{a}\rightarrow(D_{\alpha}V^{a})'=\Lambda^{a}_{b}(x)(D_{\alpha}V^{b})\label{eq2.21}
      \end{equation}
      where
      \begin{equation}
      (D_{\alpha}V^{a})'=\partial_{\alpha}V'^{a}+\omega'^{a}\medskip_{b\alpha}V'^{b}\label{eq2.22}
      \end{equation}
      
      \begin{equation}
      V'^{a}=\Lambda^{a}_{b}(x)V^{b}\label{eq2.23}
      \end{equation}
      gives rise to the following transformation rule for $ \omega_{\alpha}$
      \begin{equation}
      \Lambda:\omega^{a}\medskip_{b\alpha}\rightarrow \omega'^{a}\medskip_{b\alpha}=\Lambda^{a}_{c}\omega^{c}\medskip_{d\alpha}(\Lambda^{-1})^{d}_{b} - (\partial_{\alpha}\Lambda^{a}_{c})(\Lambda^{-1})^{c}_{b}\label{eq2.24}
      \end{equation}
The matrix $\omega_{\alpha}$ transforms under $SO(1,3)_{local}$ as a gauge field: \emph{the spin connection is the gauge field of the local Lorentz group}.
      
      If one calculates the covariant derivative of $\partial_{\alpha}$ expanded in the anholonomic basis as in (\ref{eq2.18}) using (\ref{eq2.4}), this is
      \begin{equation}
      D_{\alpha}(\partial_{\beta})=(\partial_{\alpha}e^{a}_{\beta}+\omega^{a}\medskip_{b\alpha}e^{b}_{\beta})X_{a}\label{eq2.25}
      \end{equation}
      and likewise in the holonomic basis, where
      \begin{equation}
      D_{\alpha}(\partial_{\beta})=\Gamma^{\gamma}\medskip_{\alpha\beta}\partial_{\gamma}=\Gamma^{\gamma}\medskip_{\alpha\beta}e^{a}_{\gamma}X_{a}\label{eq2.26}
      \end{equation}
      then by equality of the right hand sides one finds
      \begin{equation}
      \partial_{\alpha}e^{a}_{\beta}+\omega^{a}\medskip_{b\alpha}e^{b}_{\beta}-\Gamma^{\gamma}\medskip_{\alpha\beta}e^{a}_{\gamma}=0\label{eq2.27}
      \end{equation}
       This equation gives the relation between the spin connection $\omega$ and the $GL(4,\mathbb{R})$ connection $\Gamma$. The covariant derivative of the vierbein which has multiple indices, is obtained by using the appropriate connection term for every index. Equation (\ref{eq2.27}) thus means that the covariant derivative of the vierbein is zero
       \begin{equation}
       D_{\alpha}e^{a}_{\beta}=0\label{eq2.28}
       \end{equation}
       Therefore for every vector field $V=V^{a}X_{a}=V^{\beta}\partial_{\beta}$ we have
       \begin{equation}
       D_{\alpha}V^{a}=e^{a}_{\beta}D_{\alpha}V^{\beta}
       \end{equation}\label{eq2.29}
which shows that both versions of the covariant derivatives are equivalent.

If one calculates the covariant derivative of $g_{\alpha\beta}$
\begin{equation}
D_{\gamma}g^{\alpha\beta}=e^{\alpha}_{a}e^{\beta}_{b}D_{\gamma}\eta^{ab}=e^{\alpha}_{a}e^{\beta}_{b}(\omega^{ab}\medskip_{\gamma}+\omega^{ba}\medskip_{\gamma})\label{eq2.30}
\end{equation}
then it follows
\begin{equation}
D_{\gamma}g^{\alpha\beta}=0\Leftrightarrow\omega^{ab}=-\omega^{ba}\label{eq2.31}
\end{equation}
If $\omega^{ab}=-\omega^{ba}$, then the metric tensor and the spin connection are called compatible. A Riemannian manifold with a metric-compatible connection $\omega$, is called a \emph{Riemann-Cartan space}.
      
      From now on let $(M,g)$ be a Riemann-Cartan spacetime. The vierbein 1-form $\theta^{a}$ and the metric-compatible connection $\omega^{ab}$ are then the fundamental  geometrical variables. They satisfy the equations of Cartan
\begin{align}
\Omega^{ab}=&d\omega^{ab}+\omega^{a}\medskip_{c}\wedge \omega^{cb}=D\omega^{ab}\label{eq2.32}\\
\Theta^{a}=&d\theta^{a}+\omega^{a}\medskip_{b}\wedge \theta^{b}=D\theta^{a}\label{eq2.33}
\end{align}
with $d$ the exterior derivative and $D$ the $SO(1,3)$ exterior covariant  derivative. The two-forms $ \Omega^{ab}$  and $\Theta^{a}$ are the Riemann-Cartan curvature and torsion two-form, respectively. Using (\ref{eq2.24}) we find that under a local Lorentz transformation, the curvature form transforms as
\begin{equation}
\Omega'^{a}\medskip_{b}=\Lambda^{a}_{c}\Omega^{c}\medskip_{d}(\Lambda^{-1})^{d}_{b}\label{eq2.34}
\end{equation}
while the torsion form transforms as the co-frame
\begin{equation}
\Theta'^{a}=\Lambda^{a}_{b}\Theta^{b}\label{eq2.35}
\end{equation}
The components $R^{ab}\medskip_{\alpha\beta}$ of $\Omega^{ab}$ determined by
\begin{equation}
\Omega^{ab}=\frac{1}{2}R^{ab}\medskip_{\alpha\beta}dx^{\alpha}\wedge dx^{\beta}\label{eq2.36}
\end{equation}
and the components $T^{a}\medskip_{\alpha\beta}$ of $\Theta^{a}$ given by
\begin{equation}
\Theta^{a}=\frac{1}{2}T^{a}\medskip_{\alpha\beta}dx^{\alpha}\wedge dx^{\beta}\label{eq2.37}
\end{equation}
are called curvature tensor and torsion tensor, respectively. With (\ref{eq2.32}) and (\ref{eq2.33}) one finds their explicit form
\begin{align}
R^{ab}\medskip_{\alpha\beta}=&\partial_{\alpha}\omega^{ab}\medskip_{\beta}-\partial_\beta\omega^{ab}\medskip_\alpha+\omega^{a}\medskip_{c\alpha}\omega^{cb}\medskip_{\beta}-\omega^{a}\medskip_{c\beta}\omega^{cb}\medskip_{\alpha}\label{eq2.38}\\
T^{a}\medskip_{\alpha\beta}=&\partial_{\alpha}e^{a}_{\beta}-\partial_{\beta}e^{a}_{\alpha}+\omega^{a}\medskip_{c\alpha}e^{c}_{\beta}-\omega^{a}\medskip_{c\beta}e^{c}_{\alpha}\label{eq2.39}
\end{align}
The curvature tensor $R^{ab}\medskip_{\alpha\beta}\,$ is in fact the $SO(1,3)$ covariant field strength tensor of the $SO(1,3)$ gauge field $\omega^{ab}$. It has the same content as the curvature tensor $ R^{\alpha}\medskip_{\beta\gamma\delta}$ defined from the $ GL(4,\mathbb{R})$ connection $\Gamma^{\alpha}\medskip_{\beta}\gamma \,$ and its first derivative. They are related by
\begin{equation}
R^{a}\medskip_{b\gamma\delta}=e^{a}_{\alpha}e^{\beta}_{b}R^{\alpha}\medskip_{\beta\gamma\delta}\label{eq2.40}
\end{equation}
Analogous one has for the torsion tensor
\begin{equation}
T^{a}\medskip_{\alpha\beta}=e^{a}_{\gamma}T^{\gamma}\medskip_{\alpha\beta}\label{eq2.41}
\end{equation}
with $T^{\gamma}\medskip_{\alpha\beta}$ the expression for the torsion tensor in terms of the $GL(4,\mathbb{R})$ connection
\begin{equation}
T^{\gamma}\medskip_{\alpha\beta}=\Gamma^{\gamma}\medskip_{\alpha\beta}-\Gamma^{\gamma}\medskip_{\beta\alpha}\label{eq2.42}
\end{equation}
One can solve the expression for $T^{\gamma}\medskip_{\alpha\beta}$ given by (\ref{eq2.39}) for $\omega^{ab}\medskip_{\alpha}$, to find that 
\begin{equation}
\omega^{ab}\medskip_{\alpha}=\hat{\omega}^{ab}\medskip_{\alpha}+\phi^{ab}\medskip_{\alpha}\label{eq2.43}
\end{equation}
with $\hat{\omega}^{ab}\medskip_{\alpha}$ the torsion-free connection also called minimal or Levi-Civita connection. Explicit,
\begin{align}
\hat{\omega}^{ab}\medskip_{\alpha}=&\frac{1}{2}e^{\beta a}(\partial_{\alpha}e^{b}_{\beta}-\partial_{\beta}e^{b}_{\alpha})-\frac{1}{2}e^{\beta b}(\partial_{\alpha}e^{a}_{\beta}-\partial_{\beta}e^{a}_{\alpha})-\frac{1}{2}e^{\gamma a}e^{\beta b}(\partial_{\gamma}e_{\beta c}-\partial_{\beta}e_{\gamma c})e^{c}_{\alpha}\label{eq2.44}\\
\phi^{ab}_{\alpha}=&\frac{1}{2}e^{\beta a}T^{b}\medskip_{\beta\alpha}-\frac{1}{2}e^{\beta b}T^{a}\medskip_{\beta\alpha}-\frac{1}{2}e^{\gamma a}e^{\delta b}T^{c}\medskip_{\delta\gamma}e_{\alpha c}\label{eq2.45}
\end{align}
After substitution of (\ref{eq2.43}) into (\ref{eq2.27}) and making use of (\ref{eq2.4}) and (\ref{eq2.9}), one finds the explicit expression for the $GL(4,\mathbb{R})$ connection
\begin{equation}
\Gamma^{\alpha}\medskip_{\gamma\delta}=\lbrace_{\gamma}\medskip^{\alpha}\medskip_{\delta}\rbrace+g^{\alpha\beta}K_{\beta\gamma\delta}\label{eq2.46}
\end{equation}
with $\lbrace_{\gamma}\medskip^{\alpha}\medskip_{\delta}\rbrace$ the Christoffel symbol and
\begin{equation}
K_{\beta\gamma\delta}=\frac{1}{2}(T_{\beta\gamma\delta}-T_{\delta\beta\gamma}-T_{\gamma\beta\delta})\label{eq2.47}
\end{equation}
The expression (\ref{eq2.46}) is also found from the linear combination 
\begin{equation}
\frac{1}{2}(D_{\gamma}g_{\alpha\beta}-D_{\alpha}g_{\beta\gamma}-D_{\beta}g_{\gamma\alpha})=0\label{eq2.48}
\end{equation}
after using (\ref{eq2.42}).\\
The splitting of the Riemann-Cartan curvature two-form in a torsion-free and a torsion term, can be find by substitution of (\ref{eq2.43}) into (\ref{eq2.32})
\begin{equation}
\Omega^{ab}=(d\hat{\omega}^{ab}+\hat{\omega}^{a}\medskip_{c}\wedge\hat{\omega}^{cb})+(d\phi^{ab}+\hat{\omega}^{a}\medskip_{c}\wedge\phi^{cb}+\phi^{a}\medskip_{c}\wedge\hat{\omega}^{cb})+\phi^{a}\medskip_{c}\wedge\phi^{cb}\label{eq2.49}
\end{equation}\\[-12mm]
or
\begin{equation}
\Omega^{ab}=\hat{\Omega}^{ab}+\hat{D}\phi^{ab}+\phi^{a}\medskip_{c}\wedge\phi^{cb}\label{eq2.50}
\end{equation}
Here $\hat{\Omega}^{ab}$ and $\hat{D}$  are the curvature two-form and exterior covariant derivative calculated with the Levi-Civita connection.
\subsection{Some General Relations}
With $\lbrace \theta^{a}\rbrace$ the orthonormal co-frame, the 4-volume form $\epsilon$ is given by the Hodge dual of $1$
\begin{equation}
\epsilon =\ast 1=\frac{1}{4!}\epsilon_{abcd}\theta^{a}\wedge\theta^{b}\wedge\theta^{c}\wedge\theta^{d}=\theta^{0}\wedge\theta^{1}\wedge\theta^{2}\wedge\theta^{3}=ed^{4}x\label{eq2.51}
\end{equation}
where $\epsilon_{abcd}$ is the Levi-Civita symbol with $\epsilon_{0123}=1$, and
\begin{equation}
e=\sqrt{\vert g\vert}=det(e^{a}_{\alpha})=-\frac{1}{4!}\epsilon_{abcd}\epsilon^{\alpha\beta\gamma\delta}e^{a}_{\alpha}e^{b}_{\beta}e^{c}_{\gamma}e^{d}_{\delta}\label{eq2.52}
\end{equation}
In the coordinate basis we have the tensor density
\begin{equation}
\epsilon^{\alpha\beta\gamma\delta}=e\epsilon^{abcd}e^{\alpha}_{a}e^{\beta}_{b}e^{\gamma}_{c}e^{\delta}_{d}\label{eq2.53}
\end{equation}
Further we define
\begin{align}
\epsilon_{a}=&\ast\theta_{a}=\frac{1}{3!}\epsilon_{abcd}\theta^{b}\wedge\theta^{c}\wedge\theta^{d}=\epsilon(X_{a})\label{eq2.54}\\
\epsilon_{ab}=&\ast(\theta_{a}\wedge\theta_{b})=\frac{1}{2}\epsilon_{abcd}\theta^{c}\wedge\theta^{d}=\epsilon_{a}(X_{b})\label{eq2.55}\\
\epsilon_{abc}=&\ast(\theta_{a}\wedge\theta_{b}\wedge\theta_{c})=\epsilon_{abcd}\theta^{d}=\epsilon_{ab}(X_{c})\label{eq2.56}\\
\epsilon_{abcd}=&\ast(\theta_{a}\wedge\theta_{b}\wedge\theta_{c}\wedge\theta_{d})=\epsilon_{abc}(X_{d})\label{eq2.57}
\end{align}
These forms together with $\epsilon$ span the Grassmann algebra of exterior forms on spacetime M.\\
In the course of this work, the following relations will be of use
\begin{equation}
\epsilon=\frac{1}{4}\theta^{a}\wedge\epsilon_{a},\quad \epsilon_{a}=\frac{1}{3}\theta^{b}\wedge\epsilon_{ab},\quad \epsilon_{ab}=\frac{1}{2}\theta^{c}\wedge\epsilon_{abc},\quad \epsilon_{abc}=\theta^{d}\epsilon_{abcd}\label{eq2.58}
\end{equation}
\begin{align}
\theta^{a}\wedge\epsilon_{b}=&\delta^{a}_{b}\epsilon\label{eq2.59}\\
\theta^{a}\wedge\epsilon_{bc}=&\delta^{a}_{c}\epsilon_{b}-\delta^{a}_{b}\epsilon_{c}\label{eq2.60}\\
\theta^{a}\wedge\epsilon_{bcd}=&\delta^{a}_{d}\epsilon_{bc}+\delta^{a}_{c}\epsilon_{db}+\delta^{a}_{b}\epsilon_{cd}\label{eq2.61}\\
\theta^{a}\epsilon_{bcde}=&\delta^{a}_{e}\epsilon_{bcd}-\delta^{a}_{d}\epsilon_{bce}+\delta^{a}_{c}\epsilon_{bde}-\delta^{a}_{b}\epsilon_{cde}\label{eq2.62}\\
\epsilon^{a_{1}...a_{k}c_{k+1}...c_{4}}\epsilon_{b_{1}...b_{k}c_{k+1}...c_{4}}=&-(4-k)!\delta^{a_{1}...a_{k}}_{b_{1}...b_{k}}\label{eq2.63}
\end{align}
As a consequence of (\ref{eq2.31}), we have
\begin{equation}
D\epsilon_{abcd}=0,\quad D\epsilon_{abc}=\Theta^{d}\epsilon_{abcd},\quad D\epsilon_{ab}=\Theta^{c}\wedge\epsilon_{abc},\quad D\epsilon_{a}=\Theta^{b}\wedge\epsilon_{ab},\quad D\epsilon=0\label{eq2.64}
\end{equation}
For differential forms $\Phi $ and $\Psi $ of the same degree, we have
\begin{equation}
\ast\Phi\wedge\Psi=\ast\Psi\wedge\Phi\label{eq2.65}
\end{equation}
If $\Phi $ is a 4-form, then
\begin{equation}
\Phi(X_{a})\wedge\theta^{b}=-\delta^{b}_{a}\Phi\label{eq2.66}
\end{equation}
We will often make use of the Bianchi identities,
\begin{equation}
D\Theta^{a}=\Omega^{ab}\wedge\theta_{b},\quad D\Omega^{ab}=0\label{eq2.67}
\end{equation}
In deriving conservation laws from field equations in sections 4 and 5, the following relations will be of use
\begin{equation}
(\ast D \ast)^{2}\Omega_{ab}=0\label{eq2.68}
\end{equation}
\begin{equation}
DD\ast\Theta^{a}=\Omega^{ab}\wedge\ast\Theta_{b}\label{eq2.69}
\end{equation}
\begin{equation}
\frac{1}{2}\ast(\epsilon_{fdc[b}\Omega^{fd}\wedge\theta^{c}\wedge\theta_{a]})=\frac{1}{2}\ast(\epsilon_{abcd}\Omega^{c}\medskip_{f}\wedge\theta^{f}\wedge\theta^{d})=R^{c}_{[ab]c}\label{eq2.70}
\end{equation}
\begin{equation}
\frac{1}{2}\epsilon_{abcd}\ast(\Theta^{c}\wedge\theta^{a}\wedge\theta^{d})=2\ast(\ast\Theta_{[a}\wedge\theta_{b]}\wedge\theta^{a})=T^{a}\medskip_{ab}\label{eq2.71}
\end{equation}
\section{The Gauge Symmetries of the Lorentz Group}\label{sec.3}
\subsection{Spinor Representation}
Matter fields are represented by tensor $p$-forms $\psi $ of type $(L,E)$ with $L$ denoting the representation of the local Lorentz group in a linear space $E$. Under a local gauge transformation $\Lambda(x)\in SO(1,3) $, a section $\psi(x)$ of $E$ is thus transformed as
\begin{equation}
\psi'(x)=L(\Lambda(x))\psi(x)\label{eq3.1}
\end{equation}
For the spin indices of matter fields we have the transformation law (\ref{eq3.1}) when $L$ is a solution of
\begin{equation}
L^{-1}\gamma^{a}L=\Lambda^{a}\medskip_{b}\gamma^{b}\label{eq3.2}
\end{equation}
$L$ is the \emph{spinor representation} of the tetrad rotation $\Lambda$.\\
    
     The Dirac matrices $\gamma^{a}=\eta^{ab}\gamma_{b}$ generating the real Clifford algebra $C(1,3)$ on Minkowski space, are represented by
     \begin{equation}
     \gamma^{0}=\left(\begin{matrix}
     1&0\\0&-1 \end{matrix}\right)
     \quad,\quad \gamma^{s}=\left(\begin{matrix}
     0&\sigma_{s}\\-\sigma_{s}&0 \end{matrix}\right) \quad ,\quad  s=1,2,3 \label{eq3.3}
     \end{equation}
     with $\sigma_{s}=\sigma_{s}^{\dagger}$ the Pauli matrices providing a set of basis operators of the fundamental irrep of $su(2)$,
  \begin{equation}
     \sigma_{1}=\left(\begin{matrix}
     0&1\\1&0 \end{matrix}\right)
     \quad \sigma_{2}=\left(\begin{matrix}
     0&-i\\i&0 \end{matrix}\right)
     \quad \sigma_{3}=\left(\begin{matrix}
     1&0\\0&-1 \end{matrix}\right)\label{eq3.4}
     \end{equation} 
     The Dirac matrices satisfy
     \begin{equation}
       \frac{1}{2}(\gamma_{a}\gamma_{b}+\gamma_{b}\gamma_{a})=-\eta_{ab}\label{eq3.5}
       \end{equation}  
   The (4x4)-matrices 
   \begin{equation}
     \sigma_{ab}=-\frac{1}{4}i[\gamma_{a},\gamma_{b}]\label{eq3.6}
     \end{equation}
     provide a set of generators of the fundamental spinor irrep of $so(1,3)$. Accordingly they satisfy
     \begin{equation}
     [\sigma_{ab}, \sigma_{cd}]=i[\eta_{ad}\sigma_{bc}+\eta_{bc}\sigma_{ad}-\eta_{ac}\sigma_{bd}-\eta_{bd}\sigma_{ac}]\label{eq3.7}
     \end{equation}
     and
     \begin{equation}
     \sigma_{ab}^{\dagger}=\sigma^{ab}\label{eq3.8}
     \end{equation}
     Spacetime dependent Dirac matrices can be introduced as
     \begin{equation}
     \gamma_{\alpha}(x)=e^{a}_{\alpha}(x)\gamma_{a}\label{eq3.9}
     \end{equation}
     They satisfy
     \begin{equation}
     D_{\alpha}\gamma_{\beta}=\partial_{\alpha}\gamma_{\beta}+[\omega_{\alpha},\gamma_{\beta}]-\Gamma^{\gamma}\medskip_{\alpha\beta}\gamma_{\gamma}=0\label{eq3.10}
     \end{equation}
          where we used (\ref{eq2.27}), (\ref{eq3.5}), (\ref{eq3.6}) and the expansion of the spin connection $\omega=\omega_{\alpha}dx^{\alpha}$ in the $so(1,3)$ algebra:
          \begin{equation}
          \omega=-\frac{1}{2}i\omega^{ab}\sigma_{ab}\label{eq3.11}
          \end{equation}

The exterior covariant derivative $D\psi$ of the matter field $\psi$ transforms like a spinor
\begin{equation}
          D\psi'(x)=L(\Lambda(x))D\psi(x)\label{eq3.12}
          \end{equation} 
if it is defined as
\begin{equation}
         D\psi=d\psi+\omega\psi
         \end{equation}\label{eq3.13}
and $\omega$ transforms as
\begin{equation}
         \omega'=L\omega L^{-1}-(dL)L^{-1}\label{eq3.14}
         \end{equation} 
         
  Consider now an infinitesimal Lorentz transformation
  \begin{equation}
    \Lambda^{a}_{b}=\delta^{a}_{b}-\alpha^{a}_{b} \quad ,\quad \alpha_{ab}=-\alpha_{ba}\label{eq3.15}
    \end{equation}
    then from (\ref{eq3.2}) and the relation
    \begin{equation}
      [\sigma_{ab},\gamma_{c}]=i(\gamma_{a}\eta_{bc}-\gamma_{b}\eta_{ac)}\label{eq3.16}
      \end{equation}
 it follows that the rules for the local Lorentz transformation $\psi'=\psi+\delta\psi$ of a spinor $\psi$ are given by
 \begin{equation}
 \delta\psi=\frac{1}{2}i\alpha^{ab}\sigma_{ab}\psi\label{eq3.17}
 \end{equation}
 this is,
 \begin{equation}
 L=1+\frac{1}{2}i\alpha^{ab}\sigma_{ab}=1-\alpha \quad, \quad \alpha=-\frac{1}{2}i\alpha^{ab}\sigma_{ab}\label{eq3.18}
 \end{equation}\\
 The adjoint spinor $\bar{\psi}=\psi^{\dagger}\gamma^{0} \quad$ transforms as 
 \begin{equation}
 \bar{\psi}'(x)=\bar{\psi}(x)L^{-1}(\Lambda (x))\label{eq3.19}
 \end{equation}
 where
 \begin{equation}
 L^{-1}=1-\frac{1}{2}i\sigma_{ab}\alpha^{ab}=1+a\label{eq3.20}
 \end{equation}
 This transformation rule is derived using $\psi'^{\dagger}=\psi^{\dagger}L^{\dagger}$ , (\ref{eq3.8}) and $\sigma_{ab}\gamma^{0}=\gamma^{0}\sigma^{ab}$.
 If we define
 \begin{equation}
    D\bar{\psi}=d\bar{\psi}-\bar{\psi}\omega\label{eq3.21}
    \end{equation}   
    then $D\bar{\psi}$ transforms like $\bar{\psi}$ , this is
    \begin{equation}
    D\bar{\psi}'(x)=D\bar{\psi}(x)L^{-1}(\Lambda(x))\label{eq3.22}
    \end{equation}\\
    Under an infinitesimal Lorentz transformation (\ref{eq3.15}), the spin connection transforms as
    \begin{equation}
    \delta\omega =D\alpha=d\alpha+[\omega,\alpha]\label{eq3.23}
       \end{equation}
       This follows straightforward from $\delta\omega=\omega'-\omega$ and (\ref{eq3.14}) after substitution of (\ref{eq3.18}) and (\ref{eq3.20}). From $\delta\omega=-\frac{1}{2}i\delta\omega^{ab}\sigma_{ab}=D\alpha=-\dfrac{1}{2}D\alpha^{ab}\sigma_{ab}$ and (\ref{eq3.7}), one obtains
       \begin{equation}
       \delta \omega^{ab}=D\alpha^{ab}=d\alpha^{ab}+\omega^{ac}\alpha_{c}\medskip^{b}-\alpha^{a}\medskip_{c}\omega^{cb}\label{eq3.24}
       \end{equation}
       \subsection{Noether Identities from the Lorentz Symmetry}
       The independent variables in a SO(1,3) gauge theory are $\omega^{ab}$, $\theta^{a}$ and $\psi$. A Lagrangian describing matter fields in this gauge theory, is a functional
       \begin{equation}
       \mathcal{L} _{M}=\mathcal{L}_{M}(\theta^{a}, \psi, D\psi)\label{eq3.25}
       \end{equation}
       We say that this matter Lagrangian 4-form is locally gauge invariant if it is invariant under the simultaneous transformations (\ref{eq2.11}), (\ref{eq2.24}), (\ref{eq3.1}) and (\ref{eq3.12}). 
If $\psi$ is a spinor field, these transformations are expressed by
\begin{align}
&\delta\psi=-\alpha\psi, \quad \delta\bar{\psi}=\bar{\psi}\alpha\label{eq3.26}\\
&\delta \theta^{a}=-\alpha^{a}\medskip_{b}\theta^{b}\label{eq3.27}\\
&\delta\omega^{ab}=D\alpha^{ab}\label{eq3.28}
\end{align}
while $D\psi$ transforms as $\psi$.

     The Lagrangian for a massive fermion field in interaction with the Lorentz gauge field is
     \begin{equation}
     \mathcal{L}_{M}=-\frac{1}{2}i(\bar{\psi}\ast\gamma\wedge D\psi+D\bar{\psi}\wedge\ast \gamma\psi)-m\epsilon\bar{\psi}\psi\label{eq3.29}
     \end{equation}
     where $\gamma=\gamma^{a}\theta_{a}$  and $\ast \gamma=\gamma^{a}\epsilon_{a}$ its Hodge dual. Under an infinitesimal Lorentz transformation, we have
     \begin{align}
     \delta\epsilon=&\delta\theta^{a}\wedge\epsilon_{a}=0\label{eq3.30}\\
     \delta \epsilon_{a}=&\delta\theta^{b}\wedge\epsilon_{ab}=\alpha^{b}\medskip_{a}\epsilon_{b}\label{eq3.31}\\
      \delta \epsilon_{ab}=&\delta\theta^{c}\wedge\epsilon_{abc}=\alpha^{c}\medskip_{a}\epsilon_{cb}+\alpha^{c}\medskip_{b}\epsilon_{ac}\label{eq3.32}
      \end{align}
      Here we used equations (\ref{eq2.51}),(\ref{eq2.54})-(\ref{eq2.56}),(\ref{eq2.59})-(\ref{eq2.61}) and 
      (\ref{eq3.27}). From (\ref{eq3.31}) we have then
      \begin{equation}
      \delta(\ast \gamma)=\gamma^{a}\alpha^{b}\medskip_{a}\epsilon_{b}\label{eq3.33}
      \end{equation}
      Using this and equations (\ref{eq3.16}) and (\ref{eq3.26}) one finds that every term in the matter Lagrangian (\ref{eq3.29}) is invariant under an infinitesimal Lorentz transformation:
      \begin{equation}
      \delta\mathcal{L}_{M}=0\label{eq3.34}
      \end{equation}
      Using
      \begin{align}
      d(\bar{\psi}\alpha\ast\gamma\psi)=&D\bar{\psi}\wedge\alpha\ast\gamma\psi+\bar{\psi}\alpha D(\ast\gamma\psi)+\bar{\psi}D\alpha\wedge\ast\gamma\psi\label{eq3.35}\\
      d(\bar{\psi}\ast\gamma\alpha\psi)=&D(\bar{\psi}\ast\gamma)\alpha\psi-\bar{\psi}\ast\gamma\alpha\wedge D\psi-\bar{\psi}\ast\gamma\wedge D\alpha\psi\label{eq3.36}
      \end{align}
      the variation of $\mathcal{L}_{M}$ can be written as
      \begin{equation}
      \delta \mathcal{L}_{M}=\alpha_{ab}T^{a}\wedge\theta^{b}+\frac{1}{2}\alpha^{ab} D S_{ab}+\bar{\psi}\alpha \slashed{D}\psi-(\slashed{D} \bar{\psi})\alpha\psi\label{eq3.37}
      \end{equation}
      where
      \begin{align}
      T^{a}\wedge \theta^{b}=&-\dfrac{1}{2}i(D \bar{\psi}\gamma^{b}\psi-\bar{\psi}\gamma^{b}D\psi)\wedge\epsilon^{a}\label{eq3.38}\\   
        S_{ab}=&-\frac{1}{2}\bar{\psi}(\ast\gamma\sigma_{ab}+\sigma_{ab}\ast
\gamma)\psi\label{eq3.39}\\
     \slashed{D}\psi=&-\frac{1}{2}i(\ast\gamma\wedge D\psi-D(\ast\gamma\psi))-m\epsilon\psi\label{eq3.40}\\
     \slashed{D}\bar{\psi}=&-\frac{1}{2}i(D \bar{\psi}\wedge\ast\gamma+D(\bar{\psi}\ast\gamma))-m\epsilon\bar{\psi}\label{eq3.41}
    \end{align}
    With the equations of motion satisfied (see (\ref{eq3.45})), we have from $\delta\mathcal{L}_{M}=0$ the conservation law (Noether identity)
    \begin{equation}
    DS_{ab}=\theta_{b}\wedge T_{a}-\theta_{a}\wedge T_{b}\label{eq3.42}
    \end{equation}
    Since
    \begin{equation}
    d(\alpha^{ab}S_{ab})=D\alpha^{ab}\wedge S_{ab}+\alpha^{ab}DS_{ab}\label{eq3.43}
    \end{equation}
    we find, using (\ref{eq3.26})-(\ref{eq3.28}),
    \begin{equation}
    \delta\mathcal{L}_{M}=\delta\theta_{a}\wedge T^{a}+\frac{1}{2}\delta\omega^{ab}\wedge S_{ba}+\delta\bar{\psi}\slashed{D}\psi+\slashed{D}\bar{\psi}\delta\psi+exact form\label{eq3.44}
    \end{equation}
   which is also the expression for a general variation of $\mathcal{L}_{M}$ with respect to $(\theta^{a}, \omega^{ab}, \psi)$. Thus $ T^{a}$ is the variational derivative of $\mathcal{L}_{M}$ with respect to $\theta_{a}$ and therefore recognised as the energy-momentum 3-form. $S_{ab}$ is the variational derivative with respect to $\omega^{ab}$ and represents the spin angular-momentum 3-form. $\slashed{D}\psi$ and $\slashed{D}\bar{\psi}$  are the variational derivatives of $\mathcal{L}_{M}$ with respect to $\bar{\psi}$ and $\psi$ respectively, such that
   \begin{equation}
    \slashed{D}\psi=0, \quad \slashed{D}\bar{\psi}=0\label{eq3.45}
    \end{equation}
    are Dirac's equations of motion for the fermion field in interaction with the Lorentz gauge field. From $\bar{\psi} \slashed{D}\psi - ( \slashed{D}\bar{\psi})\psi$, one obtains the conservation law for the probability current:
   \begin{equation}
    d(\bar{\psi}\ast\gamma\psi)=0\label{eq3.46}
    \end{equation}
    Using
    \begin{equation}
     D(\ast\gamma)=\gamma^{a}D\epsilon_{a}=\gamma^{a}\Theta^{b}\wedge\epsilon_{ab}=\gamma^{a} T^{b}\medskip_{ab}\epsilon\label{eq3.47}
     \end{equation} 
     with $T^{b}\medskip_{ab}$ components of the torsion tensor in an orthonormal basis, this is $\Theta^{a}=\frac{1}{2}T^{a}\medskip_{bc}\theta^{b}\wedge \theta^{c}$, the Dirac equations (\ref{eq3.40}),(\ref{eq3.41}) can also be written as
     \begin{align}
     i\gamma^{a}(D_{a}-\frac{1}{2}T^{b}\medskip_{ba})\psi-m\psi=&0\label{eq3.48}\\
     i(D_{a}-\frac{1}{2}T^{b}\medskip_{ba})\bar{\psi}\gamma^{a}+m\bar{\psi}=&0\label{eq3.49}
     \end{align}
     where we also used that $D\psi=D_{\alpha}\psi dx^{\alpha}=e^{a}_{\alpha}D_{a}\psi dx^{\alpha}=\theta^{a}D_{a}\psi$.\\
     The energy-momentum 3-form $T_{a}$ defined through (\ref{eq3.38}) is also written as
     \begin{equation}
     T_{a}=T_{a}\medskip^{b}\epsilon_{b}=\frac{1}{2}i(D_{a}\bar{\psi}\ast\gamma\psi-\bar{\psi}\ast\gamma D_{a}\psi)\label{eq3.50}
     \end{equation}
     \begin{equation}
     T_{a}\medskip^{b}=\frac{1}{2}i(D_{a}\bar{\psi}\gamma^{b}\psi-\bar{\psi}\gamma^{b} D_{a}\psi)\label{eq3.51}
     \end{equation}
     Its trace is given by
     \begin{equation}
     T\epsilon=\frac{1}{2}i(D \bar{\psi}\wedge\ast\gamma\psi+\bar{\psi}\ast\gamma\wedge D_\psi)\label{eq3.52}
     \end{equation}
     where we used that $T_{a}\medskip^{b}\epsilon=\theta^{b}\wedge T_{a}$.\\
     If the equations of motion (\ref{eq3.45}) are satisfied, the trace becomes
     \begin{equation}
     T=-m\bar{\psi}\psi\label{eq3.53}
     \end{equation}
     The spin angular-momentum 3-form $S_{ab}$ defined through (\ref{eq3.39}) can also be written as
     \begin{equation}
     S_{ab}=\frac{1}{2}\bar{\psi}\gamma^{5}\gamma\psi\wedge\theta_{a}\wedge\theta_{b}\quad,\quad \gamma^{5}=i\gamma^{0}\gamma^{1}\gamma^{2}\gamma^{3}\label{eq3.54}
     \end{equation}
     where we used that $\lbrace\sigma_{ab},\gamma_{c}\rbrace=\epsilon_{abcd}\gamma^{5}\gamma^{d}$.
     \section{Einstein-Cartan Gravity}\label{sec.4}
     \subsection{Gauge Field Lagrangian}
     In a Lorentz gauge theory, the gauge field Lagrangian will be a functional of $\Omega^{ab}, \theta^{a} , \Theta^{a}$, i.e. it will be a 4-form,
     \begin{equation}
     \mathcal{L}=\mathcal{L}(\Omega^{ab}, \theta^{a} , \Theta^{a})\label{eq4.1}
     \end{equation}
     We say that this gauge field Lagrangian is locally gauge invariant if $\mathcal{L}$ is invariant under the simultaneous transformations (\ref{eq2.11}), (\ref{eq2.34}) and (\ref{eq3.35}). Lagrangian's with this property are for example,
     \begin{align}
     \mathcal{L}^{(\Omega)}=&-\frac{1}{2}\Omega_{ab}\wedge\ast\Omega^{ab}\label{eq4.2}\\
     \mathcal{L}^{(\Theta)}=&-\frac{1}{2}\Theta_{a}\wedge\ast\Theta^{a}\label{eq4.3}\\
     \mathcal{L}^{(EC)}=&\frac{1}{2}\Omega_{ab}\wedge\ast(\theta^{a}\wedge\theta^{b})\label{eq4.4}\\
     \mathcal{L}^{(CC)}=&\frac{1}{2}\theta_{a}\wedge\theta_{b}\wedge\ast(\theta^{a}\wedge\theta^{b})\label{eq4.5}
     \end{align}
     \subsection{Einstein-Cartan Field Equations}
     In Einstein-Cartan (EC)-gravity, the Lagrangian for a system of (minimally) coupled matter and gauge fields, is a functional
     \begin{equation}
     \mathcal{L}=\mathcal{L}(\Omega^{ab}, \theta^{a} , \psi, D\psi)=\frac{1}{8\pi l_{p}^{2}}\mathcal{L}^{EC}(\Omega^{ab},\theta^{a})+\mathcal{L}_{M}(\psi,D\psi)\label{eq4.6}
     \end{equation}
     where,
     \begin{equation}
     \mathcal{L}^{EC}=\frac{1}{4}\epsilon_{abcd}\Omega^{ab}\wedge\theta^{c}\wedge\theta^{d}=\frac{1}{2}R^{ab}_{\mu\nu}e^{\mu}_{a}e^{\nu}_{b}\epsilon\label{eq4.7}
     \end{equation}
     is the EC-Lagrangian (\ref{eq4.4}). For the matter Lagrangian we have for example the Dirac Lagrangian (\ref{eq3.29}),
     \begin{equation}
     \mathcal{L}^{D}=\lbrace\frac{1}{2}i(\bar{\psi}\gamma^{a}D_{\alpha}\psi-D_{\alpha}\bar{\psi}\gamma^{a}\psi)e^{\alpha}_{a}-m\bar{\psi}\psi\rbrace\epsilon\label{eq4.8}
     \end{equation}
     and the Lagrangian for a Yang-Mills field,
     \begin{equation}
     \mathcal{L}^{YM}=-\frac{1}{2}F\wedge\ast F=-\frac{1}{4}F_{\alpha\beta} F_{\gamma\delta} e^{\alpha}_{a} e^{\beta}_{b} e^{a\gamma}e^{b\delta}\epsilon\label{eq4.9}
     \end{equation}
     The stress-energy momentum 3-form of these matter fields is defined by
     \begin{equation}
     \delta_{\theta}\mathcal{L}_{M}=\delta\theta^{a}\wedge T_{a}\label{eq4.10}
    \end{equation}
    This gives for the Dirac Lagrangian (\ref{eq4.8}) again equation (\ref{eq3.50}) for $ T_{a}$. From (\ref{eq4.9}) we obtain,
    \begin{equation}
    T_{a}^{(YM)}=\frac{1}{2}\lbrace F(X_{a})\wedge \ast F-F\wedge(\ast F(X_{a}))\rbrace\label{eq4.11}
    \end{equation}
    where $\theta^{a}(X_{b})=\delta^{b}_{a}$. The components in $T_{a}^{(YM)}=T_{a}\medskip^{c}\epsilon_{c}$ are given by
    \begin{equation}
    T_{a}\medskip^{c}=F_{ab}  F^{cb}-\frac{1}{4}\delta^{c}_{a}F_{bd}F^{bd}
    \end{equation}\label{eq4.12}
    The spin-angular momentum 3-form which is non-zero only for fermionic matter fields is defined by
    \begin{equation}
    \delta_{\omega}\mathcal{L}_{M}=\frac{1}{2}\delta\omega^{ab}\wedge S_{ba}   
     \end{equation}\label{eq4.13}
     For $\mathcal{L}_{M}=\mathcal{L}^{(D)}$ we obtain for $S_{ab}^{(D)}$ the expression (\ref{eq3.39}) already obtained in the derivation of Noethers theorem for the Lorentz symmetry.\\
Field equations are obtained by independent variations with respect to $e^{a}_{\alpha} $ and $\omega^{ab}_{\alpha}$. In deriving these equations we discard any total divergence and use the relations
\begin{align}
\delta\theta^{a}\wedge \epsilon_{b}=&\delta e^{a}_{\alpha}e^{\alpha}_{b}\epsilon=-e^{a}_{\alpha}\delta e^{\alpha}_{b}\epsilon\label{eq4.14}\\
\delta\theta^{a}\wedge \epsilon_{a}=&\delta ed^{4}x=\delta_{\theta}\epsilon\label{eq4.15}
\end{align}
One obtains,
\begin{align}
\frac{1}{2}\epsilon_{abcd}\Omega^{ab}\wedge\theta^{c}=&8\pi l_{p}^{2}T_{d}\quad :\delta\theta\label{eq4.16}\\
\epsilon_{abcd}\Theta^{c}\wedge\theta^{d}=&8\pi l_{p}^{2}S_{ab}\;\,\, :\delta\omega\label{eq4.17}
\end{align}
\subsection{Conservation Laws}
The conservation laws in EC-gravity may be evaluated from the field equations (\ref{eq4.16}) and (\ref{eq4.17}). We first take the wedge product of (\ref{eq4.16}) with $\theta^{f}$ and change indices according to $d\leftrightarrow b$, $f\leftrightarrow a$. Taking the anti-symmetric component in the indices a and b, and the Hodge dual of the resulting equation, gives
\begin{equation}
\frac{1}{2}\ast(\epsilon_{fdc[b}\Omega^{fd}\wedge \theta^{c}\wedge\theta_{a]})=8\pi l_{p}^{2}\ast (T_{[b}\wedge\theta_{a]})\label{eq4.18}
\end{equation}
Acting with $D$ on (\ref{eq4.17}), then taking the Hodge dual, gives after using the first Bianchi identity (\ref{eq2.67}),
\begin{equation}
\frac{1}{2}\ast(\epsilon_{abcd}\Omega^{c}\medskip_{f}\wedge \theta^{f}\wedge\theta^{d})=8\pi l_{p}^{2}\ast DS_{ab}\label{eq4.19}
\end{equation}
Using (\ref{eq2.70}), we obtain by comparing (\ref{eq4.18}) and (\ref{eq4.19}), the Noether identity for the Lorentz symmetry,
\begin{equation}
\ast DS_{ab}=\ast(\theta_{b}\wedge T_{a}-\theta_{a}\wedge T_{b})\label{eq4.20}
\end{equation}
To obtain the conservation law for the energy-momentum, we take the exterior covariant derivative of the first field equation (\ref{eq4.16}),
\begin{equation}
8\pi l_{p}^{2}DT_{d}=\frac{1}{2}\epsilon_{abcd}(\Omega^{ab}\wedge\Theta^{c})\label{eq4.21}
\end{equation}
We take the interior product of this equation with $X_{f}$, and the wedge product of the resulting equation with $\theta^{d}$. After using both field equations and (\ref{eq2.66}) one obtains,
\begin{equation}
DT_{d}=\Theta^{c}(X_{d})\wedge T_{c}-\frac{1}{2}\Omega^{ab}(X_{d})\wedge S_{ab}\label{eq4.22}
\end{equation}
Since the second field equation (\ref{eq4.17}) expresses torsion via spin algebraically, the torsion vanishes identically if and only if the matter is spinless ($S_{ab}=0$). From (\ref{eq4.20}) we then have
\begin{equation}
\theta_{[a}\wedge T_{b]}=0 \quad \textrm{:\,zero torsion}\label{eq4.23}
\end{equation}
which means that $T_{ab}$ is symmetric in this case. From (\ref{eq4.22}) we then have that this symmetric energy impulse tensor is divergence free,
\begin{equation}
DT_{d}=0 \quad  \textrm{:\,zero torsion}\label{eq4.24}
\end{equation}
Finally, the first field equation (\ref{eq4.16}) reduces to Einsteins equation
\begin{equation}
R^{a}\medskip_{b}-\frac{1}{2}R\delta^{a}_{b}=\kappa T^{a}\medskip_{b} \quad  \textrm{:\,zero torsion}\label{eq4.25}
\end{equation}
Remark that the conservation laws (\ref{eq4.20}) and (\ref{eq4.22}) are the same as those resulting from local invariance under the 10-dimensional Poincaré group, of matter fields on a Riemann-Cartan spacetime \cite{bib14,bib15,bib16}. For an isolated matter system in special relativity (SR) where $\Theta^{a}=\Omega^{ab}=0$ (and the field equations (\ref{eq4.16}), (\ref{eq4.17}) are not satisfied) one obtains using Cartesian coordinates,
\begin{align}
DT_{d}=&0\label{eq4.26}\\
D(S_{ab}+x_{[a}\wedge T_{b]})=&0\label{eq4.27}
\end{align}
These equations represent the 4 plus 6 conservation laws of energy-momentum and total angular momentum (spin part + orbital part) obtainable from global Poincaré invariance of SR. They imply the existence of 4+6 conserved (=time-independent) global charges
\begin{align}
p_{a}=&\int_{H_{t}}T_{a}\label{eq4.28}\\
j_{ab}=&\int_{H_{t}}(S_{ab}+x_{[a}\wedge T_{b]})\label{eq4.29}
\end{align}
where the integrals are over all space at a fixed time and the integrand must fall rapidly enough as $\vert x \vert$ approaches infinity to obtain a finite integral. In the quantum field theory, these conserved global symmetry charges from Noethers theorem are Hermitian operators on the infinite dimensional Hilbert space of states of the field. They generate the unitary infinite dimensional representations of the Poincaré group on which Wigners mass-spin classification of elementary particles is based.
\section{Anti-de Sitter Gauge Theory}\label{sec.5}
\subsection{Gauge Field Lagangian}
Essential in the formulation of the gauge theory for gravity is the notion of fibre bundle reduction and the related concept of symmetry breaking \cite{bib17,bib18,bib19,bib20,bib21}. If $P(M,G)$ is a principal fibre bundle with structure group $G$ over space-time $M$ and $H$ a closed subgroup of $G$, then the existence of a principal subbundle $Q(M,H)$, is equivalent to the existence of a section (a ‘physical Higgs field’) $ \varphi:M\rightarrow P/H$, where $P/H$ is the associated bundle to $P$ by the action of $G$ on the coset space $G/H$. There exists a one to one correspondence between these sections and equivariant mappings $\phi:P \rightarrow G/H\subset V$ of type $(\rho,V)$ where $V$ is the vector space on which $G$ acts through a representation $\rho:G \rightarrow GL(V)$ and $G/H$ is the orbit space $\rho(G).v_{0}$, $v_{0}\in G/H$ a $H$-fixed point in $V$. In fact $Q=\phi^{-1}(v_{0})$. Using this concept of symmetry breaking, the original Lagrangian defined in the $G$-principal bundle $P$ can be re-expressed in terms of quantities defined in the $H$-principal subbundle $Q$. Therefore, let $P(M,G)$ be the anti-de Sitter frame bundle which is a $G=SO(2,3)$ principal fibre bundle, and $Q(M,H)=O(M)$ the Lorentz frame bundle, a $H=SO(1,3)$ subbundle of $P$. The anti-de Sitter algebra $ \mathfrak{g}=so(2,3)$ splits into a Lorentz subalgebra $\mathfrak{h} =so(1,3)$ and a complementary vector space $\mathfrak{t}=\mathbb{R}^{1,3}$ isomorphic with Minkowski space,
\begin{equation}
\mathfrak{g}=so(1,3)+\mathbb{R}^{1,3}\label{eq5.1}
\end{equation}
The ten generators $K_{ab}$ defining the anti-de Sitter algebra
\begin{equation}
[K_{AB},K_{CD}]=K_{AD}\eta_{BC}+K_{BC}\eta_{AD}-K_{AC}\eta_{BD}-K_{BD}\eta_{AC}\label{eq5.2}
\end{equation}
where $\eta_{AB}=diag(-1,1,1,1,-1)$, split into six generators $J_{ab}=K_{ab}$ of the Lorentz subgroup and four anti-de Sitter boosts $P_{a}=(1/l)K_{4a}$, $l$ the \emph{de Sitter length}. In the basis $\lbrace J_{ab},P_{a}\rbrace $, we have the commutation relations
\begin{align}
[J_{ab},J_{cd}]=&J_{ad}\eta_{bc}+J_{bc}\eta_{ad}-J_{ac}\eta_{bd}-J_{bd}\eta_{ac}\label{eq5.3}\\
[J_{ab},P_{c}]=&P_{a}\eta_{bc}-P_{b}\eta_{ac}\label{eq5.4}\\
[P_{a},P_{b}]=&(1/l^{2})J_{ab}\label{eq5.5}
\end{align}
If $\tilde{\mu}$ is the connection one-form of a Cartan connection on $P$, then the restriction $\mu$ of $\tilde{\mu}$ to $Q$, i.e., $\mu=\gamma^{\ast}\tilde{\mu}$ where $\gamma:Q -> P$ is the identity injection of $Q$ into its extension $P$, splits according to the Lie algebra structure into a $\mathfrak{h}$-valued part $\omega$ and a $\mathfrak{t}$-valued part $\theta$,
\begin{equation}
\mu=\omega+\theta=\frac{1}{2}\mu^{AB}K_{AB}=\frac{1}{2}\omega^{ab}J_{ab}+\theta^{c}P_{c}\label{eq5.6}
\end{equation}
where $\omega^{ab}=\mu^{ab}$ and $\theta^{c}=l\mu^{4c}$. From (\ref{eq5.4}) we have that $[\gamma_{\ast}\mathfrak{h},\mathfrak{t}]\in \mathfrak{t}$, which implies that $\omega$ is the connection one-form of a Lorentz connection on $Q$ and $\theta$ a tensor one-form of type $(AdH,\mathfrak{t})$ on $Q$, where $Ad$ denotes the adjoint representation of the symmetry group. The tensor form $\theta^{a}$ is identified as the canonical form on $O(M)$.

The reduction $\Delta=\gamma^{\ast}\tilde{\Delta}$ to $Q$ of the curvature $\tilde{\Delta}$ calculated from $\tilde{\mu}$ on $P$, can be written as
\begin{equation}
\Delta=d\mu+\frac{1}{2}[\mu,\mu]=\frac{1}{2}\Delta^{AB}K_{AB}=\Omega+\Theta+\Sigma,\quad \Delta^{AB}=d\mu^{AB}+\mu^{AC}\wedge\mu_{C}\medskip^{B}\label{eq5.7}
\end{equation}
where
\begin{align}
\Omega=&d\omega+\frac{1}{2}[\omega,\omega]=\frac{1}{2}\Omega^{ab}J_{ab},\quad \Omega^{ab}=D\omega^{ab}=d\omega^{ab}+\omega^{ac}\wedge\omega_{c}\medskip^{b}\label{eq5.8}\\
\Theta=&d\theta+[\omega,\theta]=\Theta^{a}P_{a},\quad \Theta^{a}=D\theta^{a}=d\theta^{a}+\omega^{a}\medskip_{b}\wedge\theta^{b}\label{eq5.9}\\
\Sigma=&\frac{1}{2}[\theta,\theta]=\frac{1}{2l^{2}}\theta^{a}\wedge\theta^{b}J_{ab}\label{eq5.10}
\end{align}
and $[,]$ denotes the exterior product of Lie algebra valued forms. The two-form $\Omega^{ab}$ is the curvature of the Lorentz frame bundle $O(M)$ and $\Theta^{a}$ its torsion two-form. Remark that
\begin{equation}
\Delta^{ab}=\Omega^{ab}+\frac{1}{l^{2}}\theta^{a}\wedge\theta^{b},\quad \Delta^{4c}=\frac{1}{l}\Theta^{c}\label{eq5.11}
\end{equation}
From the Bianchi identity on $P$
\begin{equation}
d\tilde{\Delta}+[\tilde{\mu},\tilde{\Delta}]=0\label{eq5.12}
\end{equation}
we find after reduction to the sub-bundle $Q$,
\begin{equation}
D\Omega=0,\quad D\Theta=[\Omega,\Theta]\label{eq5.13}
\end{equation}
or
\begin{equation}
D\Theta^{a}=\Omega^{ab}\wedge\theta_{b},\quad D\Omega^{ab}=0\label{eq5.14}
\end{equation}

Within the fibre bundle formulation of a Yang-Mills gauge theory, the gauge field Lagrangian on $P(M,G)$ is defined as a Yang-Mills Weil form \cite{bib8}.\\
\begin{equation}
L_{2}(\tilde{\Delta},\ast\tilde{\Delta})\label{eq5.15}
\end{equation}
where $L_{2}$ is an $Ad(G)$-invariant Weil polynomial of degree two on the Lie algebra $\mathfrak{g}$. In the case $\mathfrak{g}=so(2,3)$ here considered, the restriction $\Delta$ of $\tilde{\Delta}$ to $Q$ is a tensor two-form on $Q$. Then the restriction to $Q$ of the Weil form itself, i.e., $L_{2}(\Delta,\ast\Delta)=\gamma^{\ast}L_{2}(\tilde{\Delta},\ast\tilde{\Delta})$, projects to a unique four-form $\mathcal{L}$ on $M$ such that
\begin{equation}
L_{2}(\Delta,\ast\Delta)=\pi^{\ast}\mathcal{L}\label{eq5.16}
\end{equation}
where $\pi$ is the projection from $Q$ on the base $M$.

      The invariance of (\ref{eq5.15}) under general gauge transformations (the vertical automorphisms of $P$), follows from the $Ad(G)$-invariance of $L_{2}$ and the fact that $\tilde{\Delta}$  is a tensor two-form. Notice that in (\ref{eq5.16}) the four-form $\mathcal{L}$ (Lagrangian on M) is defined without using pull-backs through natural sections of $P$ defined by a choice of a local trivialisation (gauge fixing) of the principal fibre bundle.
      
      For matrix groups, an $Ad(G)$-invariant Weil polynomial of degree two on its Lie-algebra $\mathfrak{g}$ is proportional to
      \begin{equation}
      L_{2}(X,Y)=-\frac{1}{2}(tr(X)tr(Y)-tr(XY))\quad,\quad X,Y \in \mathfrak{g}\label{eq5.17}
      \end{equation}
In our case where $\mathfrak{g}=so(2,3)$, $tr(X)=0$ and (\ref{eq5.17}) is proportional to the Killing form. In the 5x5-matrix representation the generators $K_{AB}$ of the Lie algebra $so(2,3)$ are given by
\begin{equation}
(K_{AB})^{M}\medskip_{L}=\delta^{M}\medskip_{A}\eta_{BL}-\delta^{M}\medskip_{B}\eta_{AL}\label{eq5.18}
\end{equation}
It is then straightforward to obtain that
\begin{equation}
L_{2}(K_{AB},K_{CD})=C(\eta_{AC}\eta_{BD}-\eta_{AD}\eta_{BC})\label{eq5.19}
\end{equation}
and therefore
\begin{align}
&L_{2}(J_{ab},J_{cd})=C(\eta_{ac}\eta_{bd}-\eta_{ad}\eta_{bc})\label{eq5.20}\\
&L_{2}(P_{a},P_{b})=-C(1/l^{2})\eta_{ab}\label{eq5.21}\\
&L_{2}(J_{ab},P_{c})=0\label{eq5.22}
\end{align}
We introduced constant $C$, since the Weil polynomials are defined only upon an arbitrary constant. We then substitute the explicit expressions for $\Omega$, $\Theta$, $\Sigma$ as given in (\ref{eq5.7})-(\ref{eq5.10}), into the Lagrangian (\ref{eq5.15}). Using the 
multi-linearity of $L_{2}$ and (\ref{eq5.19})-(\ref{eq5.22}), we obtain a gravitational Lagrangian which besides curvature-squared and torsion-squared terms also contains the Einstein action with cosmological term. In fact,
\begin{multline}
  L_{2}(\Delta,\ast\Delta)=\frac{1}{2}C\Delta^{AB}\ast\Delta_{AB}=C(\frac{1}{2}\Omega^{ab}\wedge\ast\Omega_{ab}+\frac{1}{2l^{2}}\epsilon_{abcd}\Omega^{ab}\wedge\theta^{c}\wedge\theta^{d} \\
   +\frac{1}{4l^{4}}\epsilon_{abcd}\theta^{a}\wedge\theta^{b}\wedge\theta^{c}\wedge\theta^{d}-\frac{1}{l^{2}}\Theta^{a}\wedge\ast\Theta_{a})\label{eq5.23} 
   \end{multline} 
   If we put $C=(1/16\pi)(1/g_{l^{2}})$, $g_{l}=l_{p}/l$ where $l_{p}$ is the Planck length, then the Lagrangian defines an Einstein-Cartan theory supplemented with the curvature kinetic energy $-\frac{1}{2}\Omega^{ab}\wedge\ast\Omega_{ab}$ and the torsion kinetic energy  $-\frac{1}{2}\Theta^{a}\wedge\ast\Theta_{a}$ . One can pull-back the Lagrangian to space-time $M$ by using local sections $\sigma :U\subset M\rightarrow O(M)$. By letting,
   \begin{align}
     \sigma^{\ast}\theta^{a}=&e^{a}\medskip_{\alpha}dx^{\alpha}\label{eq5.24}\\
      \sigma^{\ast}\omega^{ab}=&\omega^{ab}\medskip_{\alpha}dx^{\alpha}\label{eq5.25}\\
       \sigma^{\ast}\Omega^{ab}=&\frac{1}{2}R^{ab}\medskip_{\alpha\beta}dx^{\alpha}\wedge dx^{\beta}\label{eq5.26}\\
    \sigma^{\ast}\Theta^{a}=&\frac{1}{2}T^{a}\medskip_{\alpha\beta}dx^{\alpha}\wedge dx^{\beta}\label{eq5.27}
  \end{align}
  we identify $\theta^{a}$ as the co-frame corresponding to an orthonormal frame $X_{a}$, $\Omega^{ab}$ and $\Theta^{a}$ as the Riemann-Cartan and torsion two-form of the base manifold M. To simplify notation we will use for these tensorial quantities on the base manifold M the same notation as their corresponding quantities on O(M).\\
  Remark that in the Lagrangian (\ref{eq5.23}), $\frac{1}{2}\Omega^{ab}\wedge\ast\Omega_{ab}$ is not a topological invariant since we work in a Riemann-Cartan space. It is only after putting torsion to zero that the connection is Levi-Civita, and this term gets proportional to the Euler form.
  \subsection{Field Equations}
  Varying the Lagrangian supplemented with the matter Lagrangian $\mathcal{L}_{M}$, with respect to $\theta^{a}$ and $\omega^{ab}$, gives the field equations 
  \begin{align}
     \frac{1}{2l^{2}}\epsilon_{abcd}\Omega^{ab}\wedge\theta^{c}+\frac{1}{l^{2}}D\ast\Theta_{d}+\frac{1}{2l^{4}}\epsilon_{abcd}\theta^{a}\wedge\theta^{b}\wedge\theta^{c}=&8\pi g_{l}^{2}T^{(M)}_{d}-\frac{1}{2}T^{(\Omega)}_{d}+\frac{1}{l^{2}}T^{(\Theta)}_{d}\label{eq5.28}\\
     \frac{1}{l^{2}}\epsilon_{abcd}\Theta^{c}\wedge\theta^{d}-\frac{2}{l^{2}}\ast\Theta_{[a}\wedge\theta_{b]}+D\ast\Omega_{ab}=&8\pi g_{l}^{2}S_{ab}\label{eq5.29}
       \end{align}
where $T^{(M)}_{d}$ is the energy-momentum 3-form of matter, $T^{(\Omega)}_{d}$ and $T^{(\Theta)}_{d}$ are the Yang-Mills energy-momentum 3-forms associated with the curvature and torsion kinetic energy respectively and where we used that,
\begin{align}
  &\delta_{\theta}\mathcal{L}_{M}=\delta\theta^{a}\wedge T_{a}^{(M)}\nonumber\\ &\delta_{\theta}(-\frac{1}{2}\Omega_{ab}\wedge\ast\Omega^{ab})=\delta\theta^{c}\wedge T_{c}^{(\Omega)}\nonumber\\
 &\delta_{\theta}(-\frac{1}{2}\Theta_{a}\wedge\ast\Theta^{a})=\delta\theta^{c}\wedge ( T_{c}^{(\Theta)}-D\ast\Theta_{c})\nonumber\\
 &\delta_{\theta}(\frac{1}{2}\epsilon_{abcd}\Omega^{ab}\wedge\theta^{c}\wedge\theta^{d})=-\delta \theta^{d}\wedge(\epsilon_{abcd}\Omega^{ab}\wedge\theta^{c})\nonumber\\
 &\delta_{\theta}(\frac{1}{4}\epsilon_{abcd}\theta^{a}\wedge\theta^{b}\wedge\theta^{c}\wedge\theta^{d})=-\delta \theta^{d}\wedge(\epsilon_{abcd}\theta^{a}\wedge\theta^{b}\wedge\theta^{c})\label{eq5.30}\\
 &\delta_{\omega}\mathcal{L}_{M}=\frac{1}{2}\delta\omega^{ab}\wedge S_{ba}\nonumber\\&\delta_{\omega}(-\frac{1}{2}\Omega_{ab}\wedge\ast\Omega^{ab})=-\delta\omega^{ab}\wedge D\ast\Omega_{ab}\nonumber\\
 &\delta_{\omega}(-\frac{1}{2}\Theta_{a}\wedge\ast\Theta^{a})=-\delta\omega^{ab}\wedge\ast\Theta_{[a}\wedge\theta_{b]}\nonumber\\
 &\delta_{\omega}(\frac{1}{2}\epsilon_{abcd}\Omega^{ab}\wedge\theta^{c}\wedge\theta^{d})=\delta\omega^{ab}\wedge\epsilon_{abcd}\Theta^{c}\wedge\theta^{d}\nonumber
\end{align}

  and,
  \begin{align}
    T_{c}^{(\Omega)}=&\frac{1}{2}\lbrace\Omega^{ab}(X_{c})\wedge\ast\Omega_{ab}-\Omega^{ab}\wedge(\ast\Omega_{ab}(X_{c}))\rbrace\label{eq5.31}\\
    T_{c}^{(\Theta)}=&\frac{1}{2}\lbrace\Theta^{a}(X_{c})\wedge\ast\Theta_{a}-\Theta^{a}\wedge(\ast\Theta_{a}(X_{c}))\rbrace\label{eq5.32}
    \end{align}
    Their components in $ T_{c}^{(\Omega)}= T_{c}\medskip^{d(\Omega)}\epsilon_{d}$ and   $ T_{c}^{(\Theta)}= T_{c}\medskip^{d(\Theta)}\epsilon_{d}$ are given by
    \begin{align}
    T_{c}\medskip^{d(\Omega)}=&R_{abcf}R^{abdf}-\frac{1}{4}\delta_{c}\medskip^{d}R_{abfg}R^{abfg}\label{eq5.33}\\
    T_{c}\medskip^{d(\Theta)}=&T^{a}\medskip_{cb}T_{a}\medskip^{db}-\frac{1}{4}\delta_{c}\medskip^{d}T^{a}\medskip_{fg}T_{a}\medskip^{fg}\label{eq5.34}
    \end{align} 
    In a coordinate basis and for a metric-compatible connection, the field equations are
    \begin{multline}
     R^{\alpha}\medskip_{d}-\frac{1}{2}R e^{\alpha}\medskip_{d}+D_{\beta}(T_{d}\medskip^{\alpha\beta})+T_{d}\medskip^{\alpha\beta}T^{\gamma}\medskip_{\beta\gamma}+\frac{1}{2}T_{d}\medskip^{\beta\gamma}T^{\alpha}\medskip_{\beta\gamma}-\frac{3}{l^{2}}e^{\alpha}\medskip_{d}\\=\kappa T_{d}\medskip^{\alpha (M)}-\frac{1}{2}l^{2}T_{d}\medskip^{\alpha(\Omega)}+T_{d}\medskip^{\alpha(\Theta)}\label{eq5.35}
     \end{multline}
 \begin{multline}
  \frac{1}{2}(e^{\alpha}\medskip_{c}T^{c}\medskip_{ab}+e^{\alpha}\medskip_{a}T^{c}\medskip_{bc}-e^{\alpha}\medskip_{b}T^{c}\medskip_{ac})+T_{[ab]}\medskip^{\alpha}+\frac{1}{2}l^{2}D_{\beta}(R_{ab}\medskip^{\alpha\beta})+R_{ab}\medskip^{\alpha\beta}T^{\gamma}\medskip_{\beta\gamma}+\frac{1}{2}R_{ab}\medskip^{\beta\gamma}T^{\alpha}\medskip_{\beta\gamma}\\=\frac{1}{2}\kappa S_{ab}\medskip^{\alpha}\label{eq5.36}
  \end{multline} 
  where $\kappa=8\pi l_{p}^{2}$. They are obtained straightforwardly by using the expressions (\ref{eq2.36}), (\ref{eq2.37}) and (\ref{eq2.6}) for $\Omega^{ab}$, $\Theta^{a}$ and $\theta^{a}$, and the general relations of section 2.3.
   \subsection{Conservation Laws}
  From the Bianchi identities and the field equations, we can derive the conservation laws for anti-de Sitter theory. We take the wedge product of the first field equation (\ref{eq5.28}) with $\theta^{f}$ and change indices according to $d\leftrightarrow b$, $f\leftrightarrow a$. Taking the anti-symmetric component in the indices $a$ and $b$, and the Hodge dual of the resulting equation gives,
  \begin{equation}
  \frac{1}{2}\ast(\epsilon_{fdc[b}\Omega^{fd}\wedge\theta^{c}\wedge\theta_{a]})+\ast(D\ast\Theta_{[b}\wedge\theta_{a]})=8\pi l_{p}^{2}\ast(T^{(M)}\medskip_{[b}\wedge\theta_{a]})\label{eq5.37}
  \end{equation}
Acting with $D$ on the second field equation (\ref{eq5.29}), then taking the Hodge dual, gives after using the first Bianchi identity and (\ref{eq2.68}),
\begin{equation}
 \frac{1}{2}\ast(\epsilon_{abcd}\Omega^{c}\medskip_{f}\wedge\theta^{f}\wedge\theta^{d})+\ast(D\ast\Theta_{[b}\wedge\theta_{a]})\\=4\pi l_{p}^{2}\ast DS_{ab}\label{eq5.38}
\end{equation}
Using (\ref{eq2.70}), we obtain by comparing (\ref{eq5.37}) and (\ref{eq5.38}), the Noether identity for the Lorentz symmetry,
\begin{equation}
\ast DS_{ab}=\ast(\theta_{b}\wedge T_{a}^{(M)}-\theta_{a}\wedge T_{b}^{(M)})\label{eq5.39}
\end{equation}
To obtain the conservation law for the energy-momentum we take the wedge product of $\theta^{a}$ with (\ref{eq5.29}). The Hodge dual of this equation gives after using (\ref{eq2.71}),
\begin{equation}
\frac{3}{2l^{2}}\epsilon_{abcd}\ast(\Theta^{c}\wedge\theta^{a}\wedge\theta^{d})+\ast(\theta^{a}\wedge D\ast\Omega_{ab})=8\pi g_{l}^{2}\ast(\theta^{a}\wedge S_{ab})\label{eq5.40}
\end{equation}
Acting with $D$ on (\ref{eq5.28}) gives after using (\ref{eq2.69}), taking the Hodge dual, and changing indices $d \leftrightarrow b$,
\begin{multline}
-\frac{3}{2l^{2}}\epsilon_{abcd}\ast(\Theta^{c}\wedge\theta^{a}\wedge\theta^{d})-\frac{1}{2}\epsilon_{abcd}\ast(\Omega^{ad}\wedge\Theta^{c})-\ast(\Omega^{a}\medskip_{b}\wedge\ast\Theta_{a})\\=8\pi l_{p}^{2}\ast DT_{b}^{(M)}-\frac{1}{2}l^{2}\ast DT_{b}^{(\Omega)}+\ast DT_{b}^{(\Theta)}\label{eq5.41}
\end{multline}
By comparing (\ref{eq5.40}) and (\ref{eq5.41}) we find,
\begin{multline}
8\pi l_{p}^{2}\ast DT_{b}^{(M)}=\frac{1}{2}l^{2}\ast DT_{b}^{(\Omega)}-\ast DT_{b}^{(\Theta)}-8\pi g_{l}^{2}\ast(\theta^{a}\wedge S_{ab})+\ast(\theta^{a}\wedge D\ast\Omega_{ab}) \\ -\ast(\Omega_{ab}\wedge\ast\Theta^{a})-\frac{1}{2}\epsilon_{abcd}\ast(\Omega^{ad}\wedge\Theta^{c})\label{eq5.42}
\end{multline}
To get this differential conservation law in the same form as (\ref{eq4.22}), we take the exterior covariant derivative of the first field equation (\ref{eq5.28}),
\begin{equation}
8\pi l_{p}^{2}\ast DT_{d}=\frac{1}{2}\epsilon_{abcd}(\Omega^{ab}\wedge\Theta^{c})+DD\ast\Theta_{d}+\frac{1}{2}l^{2} DT_{d}^{(\Omega)}-DT_{d}^{(\Theta)}\label{eq5.43}
\end{equation}
where we defined
\begin{equation}
T_{d}=T_{d}^{(M)}+T_{d}^{(V)}\quad,\quad T_{d}^{(V)}=-\frac{1}{16\pi l_{p}^{2}l^{2}}\epsilon_{abcd}\theta^{a}\wedge\theta^{b}\wedge\theta^{c}\label{eq5.44}
\end{equation}
and $T_{d}^{(V)}$ represents an anti-de Sitter cosmological term.
We take the interior product of equation (\ref{eq5.43}) with $X_{f}$, and the wedge product of the resulting equation with $\theta^{d}$. After using both field equations and (\ref{eq2.66}) one obtains,
\begin{equation}
DT_{d}=\Theta^{c}(X_{d})\wedge T_{c}-\frac{1}{2}\Omega^{ab}(X_{d})\wedge S_{ab}\label{eq5.45}
\end{equation}
We also used
\begin{align}
DT_{c}^{(\Omega)}=&D\ast\Omega_{ab}\wedge\Omega^{ab}(X_{c})-T_{d}^{(\Omega)}\wedge\Theta^{d}(X_{c})\label{eq5.46}\\
DT_{c}^{(\Theta)}=&D\ast\Theta_{a}\wedge\Theta^{a}(X_{c})-D\Theta^{a}\wedge\ast\Theta_{a}(X_{c})-T_{d}^{(\Theta)}\wedge\Theta^{d}(X_{c})\label{eq5.47}
\end{align}

\subsection{The Case of Zero Torsion}
For zero torsion, the field equations (\ref{eq5.28}-\ref{eq5.29}) reduce to,
\begin{align}
\frac{1}{2}\varepsilon_{abcd}\Omega^{ab}\wedge \theta^{c}+\frac{1}{2l^{2}}\varepsilon_{abcd}\theta^{a}\wedge \theta^{b}\wedge
\theta^{c}=&8\pi l_{p}^{2} T^{(M)}_{d} - \frac{1}{2} l^{2}T^{(\Omega)}_{d}\label{eq5.48}\\
D\ast\Omega_{ab}=&0\label{eq5.49}
\end{align}
For zero torsion one has the Levi-Civita connection, and is $D\ast=\ast D$, such that the Bianchi identity $D\Omega_{ab}=0$ implies $D\ast\Omega_{ab}=0$. From (\ref{eq5.29}) we then have that also the spin tensor $S_{ab}$ must be zero.
In component notation, these two equations are,
\begin{align}
R^{a}\medskip_{b}-\frac{1}{2}R \delta^{a}\medskip_{b}-(3/l^{2})\delta^{a}\medskip_{b}=&8\pi l_{p}^{2}T^{a}\medskip_{b}^{(M)}-\frac{1}{2}l^{2}T^{a}\medskip_{b}^{(\Omega)}\label{eq5.50}\\
D_{d}(R_{ab}\medskip^{cd})=&0\label{eq5.51}
\end{align}
where
\begin{equation}
T^{a}\medskip_{b}^{(\Omega)}=R^{cdaf}R_{cdbf}-\frac{1}{4} \delta^{a}\medskip_{b}R^{cdfg}R_{cdfg}\label{eq5.52}
\end{equation}
are the components in $ T_{a}^{(\Omega)}=T_{a}\medskip^{b(\Omega)}\varepsilon_{b} $.\\
The conservation laws (\ref{eq5.45}-\ref{eq5.46}) reduce to,
\begin{align}
DT_{a}^{(M)}=&0\label{eq5.53}\\
DT_{a}^{(\Omega)}=&0\label{eq5.54}
\end{align}
So, for zero torsion and thus zero spin current, the matter tensor and curvature kinetic energy are conserved.\\
From (\ref{eq5.52}) we see that $T^{a(\Omega)}_{\;\; b}$ is a symmetric tensor (with zero trace), implying that for zero torsion $ T^{a(M)}_{\;\; b} \;$ is also symmetric.
\subsection{Solutions}
Depending on the limit considered one obtains different theories.\\
\\
(1) $ l\rightarrow\infty \: (g_{l}\rightarrow 0) $ \\
In this limit, the theory reduces to the Stephenson-Yang theory \cite{bib22}. The Lagrangian of the theory $\frac{1}{2}\Omega^{ab}\wedge \ast\Omega_{ab}$ corresponds to that of a Yang-Mills theory of the Lorentz group. Field equations are given by
\begin{align}
T_{d}^{(\Omega)}=&0\label{eq5.55}\\
D \ast\Omega_{ab}=&0\label{eq5.56}
\end{align}
For zero torsion, every Einstein space
\begin{equation}
R_{ab}=\lambda g_{ab}\label{eq5.57}
\end{equation}
with $\lambda $ arbitrary, is a solution. By (\ref{eq5.55}) these are the only solutions. P Baekler and P B Yasskin have proved \cite{bib23} that the only spherically symmetric solutions of the torsion-free equations (\ref{eq5.55}-\ref{eq5.56}) are (i) the Schwarzschild-de Sitter metrics with arbitrary cosmological constant $\lambda$; (ii) the Nariai-Bertotti metrics with arbitrary $\lambda$; (iii) a specific family of flat spherical wave metrics. Remark that the cosmological constant $\lambda$ is not a fundamental constant as in Einsteins theory (there is no cosmological term in the Lagrangian in this limit). Rather, $\lambda$ is a property of the solution (an integration constant). Our theory has no matter solutions in the limit here considered. So, $l\rightarrow \infty $ corresponding to an anti-de Sitter $\rightarrow$ Poincaré contraction of the spacetime symmetry group, also corresponds to the "vacuum limit". The original Stephenson-Yang theory has serious difficulties with matter solutions and the field equations do not reduce to Poisson's equation in the Newtonian limit.\\
\\
(2) $ l \rightarrow\l_{p} \: (g_{l}\rightarrow 1) $ \\
The field equations (\ref{eq5.50}-\ref{eq5.51}) become,
\begin{align}
R^{a}\medskip_{b}-\frac{1}{2}R \delta^{a}\medskip_{b}-(3/l_{p}^{2})\delta^{a}\medskip_{b}=&8\pi l_{p}^{2}T^{a}\medskip_{b}^{(M)}-\frac{1}{2}l_{p}^{2}T^{a}\medskip_{b}^{(\Omega)}\label{eq5.58}\\
D_{d}(R_{ab}\medskip^{cd})=&0\label{eq5.59}
\end{align}
The effective matter Lagrangian must contain a counter term for the anti-de Sitter  cosmological term and contributions from the Standard Model of elementary particle physics (which are of the same order and sign as the anti-de Sitter term). This is, a 'bare' cosmological term must be fine tuned with these cosmological terms to obtain the observed cosmological constant $\Lambda_{obs}$. This is equivalent to the so called cosmological constant problem from General Relativity. The effective (renormalised) Einstein field equation then becomes
\begin{equation}
R^{a}\medskip_{b}-\frac{1}{2}R \delta^{a}\medskip_{b}+\Lambda_{obs}\delta^{a}\medskip_{b}=8\pi l_{p}^{2}T^{a}\medskip_{b}^{(M)}-\frac{1}{2}l_{p}^{2}T^{a}\medskip_{b}^{(\Omega)}\label{eq5.60}
\end{equation}\\[-10mm]
and where $T^{a(M)}_{\;b}$ contains only contributions from ordinary and dark matter. If the curvature $<< 1/l_{p}^{-2}$, the Yang-Mills curvature energy-momentum term $T^{a}\medskip_{b}^{(\Omega)}$ can be neglected and the field equation (\ref{eq5.60}) reduces to Einstein's equation.

The Schwarzschild solution in vacuum, this is for:
\begin{equation}
 T_{ab}^{(M)}=S_{ab}\medskip^{c}=\Lambda_{obs}=0\label{eq5.59}
 \end{equation}
exists only if also $T^{a(\Omega)}_{\;\; b}=0 $.  An explicit calculation of this result is given in App.\ref{App.A}. In this context see also ref.\cite{bib24}.
 
Inside matter, the Yang-Mills  field equation (5.61) imposes restrictions on the solution set of the first field equations. For example for the RW-geometry, this Yang-Mills equation determines the equation of state for the matter present. The only solution found for a RW-geometry with a perfect fluid matter content is the early universe. This is shown in App.\ref{App.B}.
  
      Solutions of gravitational theories based on a gauge invariant field Lagrangian at most quadratic in curvature and torsion (as is (\ref{eq5.23})) have also been studied by H Goenner and F Müller-Hoissen \cite{bib25}, and by J.Garecki \cite{bib26}.\\[-10mm]
\section{Discussion}\label{sec.6}
It is straightforward to translate equations obtained here for the anti-de Sitter gauge theory to the de Sitter gauge theory for gravity. In fact since the need for renormalizing the cosmological constant, conclusions will be the same. Although there is a unique embedding of $SO(2,3)$ into the conformal symmetry $SO(2,4)$ of 4D spacetime, this is not the case for $SO(1,4)$. Since there is no mass or length scale in a universe having conformal symmetry, the extension of the Standard Model symmetries with the 4D conformal symmetry is not forbidden by the Coleman-Mandula theorem. In fact, it is the only non-supersymmetric extension of the Standard Model symmetries that is not in conflict with the theorem. It is only after symmetry breaking of the conformal symmetry along the unique sequence $SO(2,4)\supset SO(2,3)\supset SO(1,3)$ that interacting Lagrangian theories can be defined, especially quantum field theories based on the Poincaré and unitary symmetries. So, in the context of unification of all fundamental forces, using the anti-de Sitter symmetry is more promising. These ideas will be explored in future work.

In this work, the gauge field Lagrangian is of the Yang-Mill type. To have a correct formulation of it in this non-unitary context, it is essential to use the fibre bundle formalism and define the Yang-Mills Lagrangian as an $Ad(G)$-invariant Weil form as was done in section 5. Applying independent variations  to it with respect to the vierbein and spin connections, results in an Einstein- and Yang-Mills-type equation which extend the field equations obtained in the Einstein-Cartan theory with extra torsion terms. The Yang-Mills equation is now dynamical and is for zero torsion the field equation for an $SO(1,3)$ Yang-Mills gauge theory. Also for zero torsion the Einstein-like equation is identical to Einstein's equation supplemented with 
a curvature kinetic energy term. This curvature kinetic energy is for zero torsion and spin current, together with the matter energy, a conserved quantity in the gauge theory.

In App.\ref{App.A} it was shown that Schwarzschild vacuum solutions are possible only for weak gravitational fields where the curvature squared terms of the curvature kinetic energy can be neglected. In strong gravitational fields as in the center of the galaxy, it is then expected that the curvature kinetic energy will cause deviations from Schwarzschild geometry geodesic motion of orbiting stars.  Notice that the physics governing the behaviour of the masses at the center of our galaxy is still under discussion. For example, the central black hole model for $Sgr A^{\ast}$ has recently been challenged. It was shown \cite{bib27} that astronomy data of orbiting stars and gaseous clouds are better fitted by geodesics in the spacetime of a self gravitating dense core of dark fermions. A subject for further study is then to use the anti de Sitter field equations for gravity in this context.
Further, the result that a FRW-geometry has the early universe as the unique solution, implicates that a gauge theory for gravity as considered here, must be seen as a possible fundamental theory.

\newpage  
\begin{appendices}
\addcontentsline{toc}{section}{Appendices}
\section*{\textbf{\Large Appendices}}
In these appendices we study solutions of the torsion free field equations for a spherical symmetric (Schwarzschild) geometry and for the homogeneous and isotropic FRW-geometry. For the computation of curvature and other geometric quantities using exterior differential forms see for example ref.\cite{bib28}-chapter 14.
 
\section{Schwarzschild Geometry}\label{App.A}

\subsection{Geometrical Quantities}
The Schwarzschild metric,
\begin{equation}
ds^{2}=-e^{2\lambda}dt^{2}+e^{2\mu}dr^{2}+r^{2}(d\theta^{2}+sin^{2}\theta d\phi^{2})
\end{equation}
with $ \lambda=\lambda(r) , \mu=\mu(r) $.
\\In a non-holonomic basis,
\begin{align}
ds^{2}=&\eta_{ab}\theta^{a}\wedge\theta^{b}\\
\theta^{t}=&e^{\lambda}dt\\
\theta^{r}=&e^{\mu}dr\\
\theta^{\theta}=&rd\theta\\
\theta^{\phi}=&rsin\theta d\phi
\end{align}
Calculate $ d\theta^{a}=-\omega^{a}\medskip_{b}\wedge\theta^{b} $ and read the connection one-forms,
\begin{align}
\omega^{t}\medskip_{r}=&\lambda^{'}e^{-\mu}\theta^{t}=\lambda^{'}e^{\lambda-\mu}dt\\
\omega^{\theta}\medskip_{r}=&\dfrac{1}{r}e^{-\mu}\theta^{\theta} =e^{-\mu} d\theta\\
\omega^{\phi}\medskip_{r}=&\dfrac{1}{r}e^{-\mu} \theta^{\phi} =e^{-\mu}sin\theta d\phi\\
\omega^{\phi}\medskip_{\theta}=&\dfrac{1}{r}cotg\theta  \theta^{\phi} =cos\theta d\phi
\end{align}
Then calculate the $\Omega^{a}_{\;b}= d\omega^{a}_{\;b}+\omega^{a}_{\;c}\wedge\omega^{c}_{\;b}$ and read the Riemann-tensor components from $\Omega^{a}_{\;b}=\dfrac{1}{2}R^{a}_{\;bcd}\theta^{c}\wedge\theta^{d}$.
\begin{align}
\Omega^{t}\medskip_{r}=&-A\theta^{t}\wedge \theta^{r}  ,  A=e^{-2\mu}(\lambda^{''}+\lambda^{'2}-\lambda^{'}\mu^{'})=-R^{t}\medskip_{rtr}\label{eqA.11}\\
\Omega^{t}\medskip_{\theta}=&-B\theta^{t}\wedge \theta^{\vartheta}  , B=\dfrac{\lambda^{'}}{r}e^{-2\mu} =-R^{t}\medskip_{\theta t \theta}\label{eqA.12}\\
\Omega^{t}\medskip_{\phi}=&-B\theta^{t}\wedge \theta^{\phi},  B=\dfrac{\lambda^{'}}{r}e^{-2\mu} =-R^{t}\medskip_{\phi t \phi}\label{eqA.13}\\
\Omega^{r}\medskip_{\theta}=&\quad C\theta^{r}\wedge \theta^{\theta} , C=\dfrac{\mu^{'}}{r}e^{-2\mu}=R^{r}\medskip_{\theta r \theta}\label{eqA.14}\\
\Omega^{r}\medskip_{\phi}=&\quad C\theta^{r}\wedge \theta^{\phi} , C=\dfrac{\mu^{'}}{r}e^{-2\mu}=R^{r}\medskip_{\phi r \phi}\label{eqA.15}\\
\Omega^{\theta}\medskip_{\phi}=&\quad D\theta^{\theta}\wedge \theta^{\phi} , D=\dfrac{1}{r^{2}}(1-e^{-2\mu})=R^{\theta}\medskip_{\phi \theta\phi}\label{eqA.16} 
\end{align}
Obtain the components of the Ricci-tensor from $ R_{ab}=R^{c}\medskip_{acb} $ and the Ricci-scalar $ R=R^{a}\medskip_{b} $ 
\begin{align}
R^{t}\medskip_{t}=&-e^{-2\mu}(\lambda^{''}+\lambda^{'2}-\lambda^{'}\mu^{'}+2\dfrac{\lambda^{'}}{r})=-A-2B\\
R^{r}\medskip_{r}=&-e^{-2\mu}(\lambda^{''}+\lambda^{'2}-\lambda^{'}\mu^{'}-2\dfrac{\mu^{'}}{r})=-A+2C\\
R^{\theta}\medskip_{\theta}=&R^{\phi}\medskip_{\phi}=e^{-2\mu}(\dfrac{\mu^{'}}{r}-\dfrac{\lambda^{'}}{r}-\dfrac{1}{r})+\dfrac{1}{r^{2}}=-B+C+D\\
R=&e^{-2\mu}(-2\lambda^{''}-2\lambda^{'2}+2\lambda^{'}\mu^{'}-4\dfrac{\lambda^{'}}{r}+4\dfrac{\mu^{'}}{r}-\dfrac{2}{r^{2}})+\dfrac{2}{r^{2}}=-2A-4B+4C+2D
\end{align}
The geometrical quantities obtained so far allow to calculate the Einstein field equations. Here however, to calculate the Yang-Mills energy-momentum 3-form associated with the curvature kinetic energy $ T_{a}^{\;(\Omega)} $ and the Yang-Mills field equations, we need more geometrical quantities such as the contractions of the curvature 2-form and its dual with the vector fields $ X_{a}$ dual to the basis 1-forms $ \theta^{a} $.\\
The contractions  $\Omega^{ab}(X_{c})$  are obtained using $\theta^{a}(X_{b})=\delta^{a}\medskip_{b} $,
\begin{equation}
\begingroup
\setlength\arraycolsep{5pt}
\begin{pmatrix}
\Omega^{tr}(X_{t})=-A\theta^{r} &
\Omega^{tr}(X_{r})=A\theta^{t} &
\Omega^{tr}(X_{\theta})=0 &
\Omega^{tr}(X_{\phi})=0 \\
\Omega^{t\theta}(X_{t})=-B\theta^{\theta} &
\Omega^{t\theta}(X_{r})=0 &
\Omega^{t\theta}(X_{\theta})=B\theta^{t} &
\Omega^{t\theta}(X_{\phi})=0 \\
\Omega^{t\phi}(X_{t})=-B\theta^{\phi} &
\Omega^{t\phi}(X_{r})=0 &
\Omega^{t\phi}(X_{\theta})=0 &
\Omega^{t\phi}(X_{\phi})=B\theta^{t}\\
\Omega^{r\theta}(X_{t})=0 &
\Omega^{r\theta}(X_{r})=C\theta^{\theta} &
\Omega^{r\theta}(X_{\theta})=-C\theta^{r} &
\Omega^{r\theta}(X_{\phi})=0 \\
\Omega^{r\phi}(X_{t})=0 &
\Omega^{r\phi}(X_{r})=C\theta^{\phi} &
\Omega^{r\phi}(X_{\theta})=0 &
\Omega^{r\phi}(X_{\phi})=-C\theta^{r} \\
\Omega^{\theta\phi}(X_{t})=0 &
\Omega^{\theta\phi}(X_{r})=0 &
\Omega^{\theta\phi}(X_{\theta})=D\theta^{\phi} &
\Omega^{\theta\phi}(X_{\phi})=-D\theta^{\theta} \\ 
\end{pmatrix}
\endgroup
\end{equation}
The dual curvature$\; \ast\Omega^{ab} $ , is obtained using $ \;\varepsilon_{ab}=\ast(\theta_{a}\wedge\theta_{b})=\frac{1}{2}\varepsilon_{abcd}\theta^{c}\wedge\theta^{d} $.
\begin{align}
\ast\Omega^{tr}=&A\varepsilon_{tr},\;
\varepsilon_{tr}=\varepsilon_{tr\theta\phi}\theta^{\theta}\wedge\theta^{\phi}\\
\ast\Omega^{t\theta}=&B\varepsilon_{t\theta},\; 
\varepsilon_{t\theta}=\varepsilon_{t\theta r\phi}\theta^{r}\wedge\theta^{\phi}\\
\ast\Omega^{t\phi}=&B\varepsilon_{t\phi},\; 
\varepsilon_{t\phi}=\varepsilon_{t\phi r\theta}\theta^{r}\wedge\theta^{\phi}\\
\ast\Omega^{r\theta}=&C\varepsilon_{r\theta},\; 
\varepsilon_{r\theta}=\varepsilon_{r\theta t\phi}\theta^{t}\wedge\theta^{\phi}\\
\ast\Omega^{r\phi}=&C\varepsilon_{r\phi},\;
\varepsilon_{r\phi}=\varepsilon_{r\phi t\theta}\theta^{t}\wedge\theta^{\theta}\\
\ast\Omega^{\theta\phi}=&D\varepsilon_{\theta\phi},\; 
\varepsilon_{\theta\phi}=\varepsilon_{\theta\phi tr}\theta^{t}\wedge\theta^{r}
\end{align}
\\ 
The contractions $ \ast\Omega^{ab}(X_{c}) $,  are obtained using $ \varepsilon_{ab}(X_{c})=\varepsilon_{abc}=\varepsilon_{abcd}\theta^{d} $,
\begin{equation}
\begingroup
\setlength\arraycolsep{5pt}
\begin{pmatrix}
\ast\Omega^{tr}(X_{t})=0 &
\ast\Omega^{tr}(X_{r})=0 &
\ast\Omega^{tr}(X_{\theta})=A\varepsilon_{tr\theta} &
\ast\Omega^{tr}(X_{\phi})=A\varepsilon_{tr\phi}\\
\ast\Omega^{t\theta}(X_{t})=0 &
\;\ast\Omega^{t\theta}(X_{r})=B\varepsilon_{t\theta r} &
\;\ast\Omega^{t\theta}(X_{\theta})=0 &
\;\ast\Omega^{t\theta}(X_{\phi})=B\varepsilon_{t\theta\phi}\\
\ast\Omega^{t\phi}(X_{t})=0 &
\ast\Omega^{t\phi}(X_{r})=B\varepsilon_{t\phi r} &
\ast\Omega^{t\phi}(X_{\theta})=B\varepsilon_{t\phi\theta} &
\ast\Omega^{t\phi}(X_{\phi})=0\\
\ast\Omega^{r\theta}(X_{t})=C\varepsilon_{r\theta t} &
\ast\Omega^{r\theta}(X_{r})=0 &
\ast\Omega^{r\theta}(X_{\theta})=0 &
\ast\Omega^{r\theta}(X_{\phi})=C\varepsilon_{r\theta\phi} \\
\ast\Omega^{r\phi}(X_{t})=C\varepsilon_{r\phi t} &
\ast\Omega^{r\phi}(X_{r})=0 &
\ast\Omega^{r\phi}(X_{\theta})=C\varepsilon_{r\phi\theta} &
\ast\Omega^{r\phi}(X_{\phi})=0 \\
\ast\Omega^{\theta\phi}(X_{t})=D\varepsilon_{\theta\phi t} &
\ast\Omega^{\theta\phi}(X_{r})=D\varepsilon_{\theta\phi r} &
\ast\Omega^{\theta\phi}(X_{\theta})=0 &
\ast\Omega^{\theta\phi}(X_{\phi})=0\\
\end{pmatrix}
\endgroup
\end{equation}
We now have all the components to calculate the Yang-Mills energy-momentum 3-form of the gravitational field:
\begin{equation}
T_{c}\medskip^{(\Omega)}=\frac{1}{2}\lbrace \Omega^{ab}(X_{c})\wedge\ast\Omega_{ab} - \Omega^{ab}\wedge\ast\Omega_{ab}(X_{c})\rbrace
\end{equation}
Use relations as: $ \theta^{a}\wedge\varepsilon_{bc}=\delta^{a}_{\;c}\varepsilon_{b} - \delta^{a}_{\;b}\varepsilon_{c}\; $ and $\;\varepsilon_{a}=\frac{1}{3!}\varepsilon_{abcd}\theta^{b}\wedge\theta^{c}\wedge\theta^{d} $ to find,
\begin{align}
T_{t}\medskip^{(\Omega)}=&(A^{2}+2B^{2}-2C^{2}-D^{2})\varepsilon_{t}=T_{t}\medskip^{ t (\Omega)}\varepsilon_{t}\\
T_{r}\medskip^{(\Omega)}=&(A^{2}-2B^{2}+2C^{2}-D^{2})\varepsilon_{r}=T_{r}\medskip^{ r (\Omega)}\varepsilon_{r}\\
T_{\theta}\medskip^{(\Omega)}=&(-A^{2}+D^{2})\varepsilon_{\theta}=T_{\theta}\medskip^{ \theta (\Omega)}\varepsilon_{\theta}\\
T_{\phi}\medskip^{(\Omega)}=&(-A^{2}+D^{2})\varepsilon_{\phi}=T_{\phi}\medskip^{ \phi (\Omega)}\varepsilon_{\phi}
\end{align}
where,
\begin{align}
\varepsilon_{t}=&\varepsilon_{tr\theta\phi}\theta^{r}\wedge\theta^{\theta}\wedge\theta^{\phi}=\theta^{r}\wedge\theta^{\theta}\wedge\theta^{\phi}\\
\varepsilon_{r}=&\varepsilon_{rt\theta\phi}\theta^{t}\wedge\theta^{\theta}\wedge\theta^{\phi}=-\theta^{t}\wedge\theta^{\theta}\wedge\theta^{\phi}\\
\varepsilon_{\theta}=&\varepsilon_{\theta t r\phi}\theta^{t}\wedge\theta^{r}\wedge\theta^{\phi}=\theta^{t}\wedge\theta^{r}\wedge\theta^{\phi}\\
\varepsilon_{\phi}=&\varepsilon_{\phi t r \theta}\theta^{t}\wedge\theta^{r}\wedge\theta^{\theta}=-\theta^{t}\wedge\theta^{r}\wedge\theta^{\theta}
\end{align}
Next we calculate the exterior covariant derivative of the dual curvature:
\begin{equation}
D\ast\Omega^{a}\medskip_{b}=d \ast\Omega^{a}\medskip_{b}+\omega^{a}\medskip_{c}\wedge \ast\Omega^{c}\medskip_{b}- \Omega^{a}\medskip_{c}\wedge \ast\omega^{c}\medskip_{b}
\end{equation}\\[-16mm]
The non-zero components are,
\begin{align}
D\ast\Omega^{t}\medskip_{r}=&e^{-\mu}(A^{'}+\frac{2}{r}(A-B))\varepsilon_{t}\\
D\ast\Omega^{r}\medskip_{\theta}=&e^{-\mu}(C^{'}+\lambda^{'}(B+C)+\frac{1}{r}(C-D))\varepsilon_{\theta}\\
D\ast\Omega^{r}\medskip_{\phi}=&e^{-\mu}(C^{'}+\lambda^{'}(B+C)+\frac{1}{r}(C-D))\varepsilon_{\phi}
\end{align}
Finally the Bianchi-identities,
\begin{equation}
D\Omega^{a}\medskip_{b}=d\Omega^{a}\medskip_{b}+\omega^{a}\medskip_{c}\wedge\Omega^{c}\medskip_{b}- \Omega^{a}\medskip_{c}\wedge\omega^{c}\medskip_{b}=0
\end{equation}\\[-20mm]
give two independent equations,
\begin{align}
B^{'}+(B+C)\lambda^{'}+\dfrac{1}{r}(B-A)=&0\label{eqA.43}\\
D^{'}+\frac{2}{r}(D-C)=&0\label{eqA.44}
\end{align}
\\[-4mm]
\subsection{Field Equations}
We consider vacuum solutions with dark energy (Schwarzschild-de Sitter spaces). Thus,
\begin{equation}
T_{ab}^{(M)}=0, \qquad S_{ab}=0, \qquad \Lambda=\Lambda_{obs}\label{eqA.45}
\end{equation}
We will not assume a priori the limit $l\rightarrow l_{p}$. 
Then the Einstein equation (\ref{eq5.60}) and Yang-Mills equation (\ref{eq5.49}) become
\begin{align}
R^{a}\medskip_{b}-\frac{1}{2}R\delta^{a}\medskip_{b}+\Lambda \delta^{a}\medskip_{b}=&-\frac{1}{2}l^{2}T^{a}\medskip_{b}^{(\Omega)}\label{eqA.46}\\
D\ast\Omega_{ab}=&0\label{eqA.47}
\end{align}
This gives tree Einstein equations: the tt-, rr-, and $ \theta\theta=\phi\phi $ -component,
\begin{align}
A+2B+\Lambda=&\frac{1}{2}l^{2}(A^{2}+2B^{2}-2C^{2}-D^{2})\label{eqA.48}\\
A-2C+\Lambda=&\frac{1}{2}l^{2}(A^{2}-2B^{2}+2C^{2}-D^{2})\label{eqA.49}\\
B-C-D+\Lambda=&\frac{1}{2}l^{2}(-A^{2}+D^{2})\label{eqA.50}
\end{align}
and two Yang-Mills equations: the tr- and $ r\theta=r\phi $ -component,
\begin{align}
A^{'}+\frac{2}{r}(A-B)=&0\label{eqA.51}\\
C^{'}+(B+C)\lambda^{'}+\frac{1}{r}(C-D))=&0\label{eqA.52}
\end{align}
Remark that,
\begin{equation}
R=-4\Lambda=2(A-D)+4(B-C)\label{eqA.53}
\end{equation}
\subsection{Solutions}
\subsubsection{Schwarzschild black-hole}
Here dark energy and $ R^{2} $-terms are neglected in the field equations. The three Einstein equations are then,
\begin{align}
A+2B=&0\label{eqA.54}\\
A-2C=&0\label{eqA.55}\\
B-C-D=&0\label{eqA.56}
\end{align}
and the Yang-Mills equations reduce to
\begin{align}
A^{'}=&-\frac{2}{r}(A-B)\label{eqA.57}\\
C^{'}=&-\lambda^{'}(B+C)-\frac{1}{r}(C-D)\label{eqA.58}
\end{align}
From (\ref{eqA.54}-\ref{eqA.56}) one obtains,
\begin{equation}
B=-C=\frac{1}{2}D=-\frac{1}{2}A\label{eqA.59}
\end{equation}
Substitution in the Yang-Mills equations (\ref{eqA.57}-\ref{eqA.58}) reduces these to,
\begin{equation}
B^{'}=-\frac{3}{r}B\label{eqA.60}
\end{equation}
From $ B=-C $ we have
\begin{equation}
\lambda^{'}+\mu^{'}=0
\end{equation}
and with $ k $ as integration constant:
\begin{equation}
\lambda+\mu=k
\end{equation}
We can chose the constant $ k $ to be zero by replacing the time coordinate t by another coordinate $ te^{-k} $ . This is equivalent to replacing $ \lambda $ by $ \lambda-k $, so that
\begin{equation}
\lambda+\mu=0
\end{equation}
Since $ B=\frac{\lambda^{'}}{r}e^{-2\mu} \;$, the Yang-Mills equation (\ref{eqA.60}) is explicitly:
\begin{equation}
\lambda ^{''}-2\lambda^{'}\mu^{'}+\frac{2}{r}\lambda^{'}=0
\end{equation}
Compare this to the Einstein equation $ A+2B=0 $  
\begin{equation}
\lambda ^{''}+\lambda^{'2}-\lambda^{'}\mu^{'}+\frac{2}{r}\lambda^{'}=0
\end{equation}
to see that with $ \lambda^{'}=-\mu^{'} $ both equations are equivalent.
So here, the solution of the Einstein equations also satisfies the Yang-Mills equation.\\
Finally, from $ B-C-D=0 \;$ and $ \lambda^{'}=-\mu^{'} \;$ we have,
\begin{equation}
(e^{-2\mu}r)^{'}=1
\end{equation}
Integration with $ -2GM \; $ as integration constant gives,
\begin{equation}
e^{-2\mu}=1-\frac{2GM}{r}=1-\frac{r_{s}}{r} \qquad (G=l_{p}^{2})
\end{equation}
were $ r=r_{s}=2GM \; $ is the \emph{event-horizon}.
\subsubsection{Schwarzschild-de Sitter space}
Here only the $ R^{2} $-terms are neglected in the field equations. The Einstein equations give,
\begin{align}
A+2B+\Lambda=&0\label{eqA.68}\\
A-2C+\Lambda=&0\label{eqA.69}\\
B-C-D+\Lambda=&0\label{eqA.70}
\end{align}
and the Yang-Mills equations,
\begin{align}
A^{'}=&-\frac{2}{r}(A-B)\label{eqA.71}\\
C^{'}=&-\lambda^{'}(B+C)-\frac{1}{r}(C-D)\label{eqA.72}
\end{align}
The Einstein-equations reduce to,
\begin{equation}
B=-C=\frac{1}{2}(D-\Lambda)=-\frac{1}{2}(A+\Lambda)\label{eqA.73}
\end{equation}
Substitution in the Yang-Mills equations gives then,
\begin{equation}
B^{'}=-\frac{3}{r}(B+\frac{1}{3}\Lambda)\label{eqA.74}
\end{equation}
From $ B=-C $ we still have that
\begin{equation}
\lambda+\mu=0
\end{equation}
Since $ B=\frac{\lambda^{'}}{r}e^{-2\mu} \;$, the Yang-Mills equation (A74) is explicitly:
\begin{equation}
\lambda ^{''}-2\lambda^{'}\mu^{'}+\frac{2}{r}\lambda^{'}=-\Lambda e^{2\mu}
\end{equation}
Also now, this equation is equivalent to the Einstein equation $ A+2B+\Lambda=0 $ with $ \lambda^{'}=-\mu^{'} $. 
So again, the solution to the Einstein equations also satisfies the Yang-Mills equation.\\
Finally, from $ B-C-D+\Lambda=0 \;$ and $ \lambda^{'}=-\mu^{'} \;$ we have,
\begin{equation}
(e^{-2\mu}r+\frac{1}{3}\Lambda r^{3})^{'}=1
\end{equation}
Integration with $ -2GM \; $ as integration constant gives,
\begin{equation}
e^{-2\mu}=1-\frac{2GM}{r}=1-\frac{r_{s}}{r}-\frac{1}{3}\Lambda r^{2} \qquad (G=l_{p}^{2})
\end{equation}
For $ r\gg r_{s}\; $ we have de Sitter space. For $ r\ll (\frac{r_{s}}{\Lambda})^{\frac{1}{3}} \; $ we recover the Schwarzschild solution.
\subsubsection{General case}
First we write the Einstein field equations (\ref{eqA.48}-\ref{eqA.50}) as,
\begin{align}
(B+C)(1-l^{2}(B-C))=&0\label{eqA.79}\\
(A-D)+2(B-C+\Lambda)=&0\label{eqA.80}\\
(A+D)(1+2l^{2}(B-C+\Lambda))=&0\label{eqA.81}
\end{align}
The first equation is obtained from (\ref{eqA.48}) minus (\ref{eqA.49}). The second from (\ref{eqA.48}) plus (\ref{eqA.49}) plus 2-times (\ref{eqA.50}). The third equation from (\ref{eqA.49}) minus (\ref{eqA.50}) and substitution of (\ref{eqA.79}) and (\ref{eqA.80}).\\
There are three cases for which these Einstein equations are satisfied:\\\\
\textbf{\emph{Case 1}}
\begin{equation}
B=-C, \quad A=-D\label{eqA.82}
\end{equation}
Then equations (\ref{eqA.79}) and (\ref{eqA.81}) are satisfied and (\ref{eqA.80}) reduces to
\begin{equation}
B=-\dfrac{1}{2}(A+\Lambda)\label{eqA.83}
\end{equation}
But this is just one of the Einstein equations (\ref{eqA.73}) where we had put $ T^{a\;(\Omega)}_{b}=0 $, a trivial result now since $ A^{2}=D^{2}\; and \;B^{2}=C^{2} $.  
So we find the same solutions as before with the Yang-Mills equations identically satisfied.\\\\
\textbf{\emph{Case 2}}
\begin{equation}
B-C=\frac{1}{l^{2}}\label{eqA.84}
\end{equation}
Then (\ref{eqA.79}) is satisfied and (\ref{eqA.81}) reduces to
\begin{equation}
(A+D)(3+2l^{2}\Lambda)=0\label{eqA.85}
\end{equation}
implying $ A=-D\; $. So, in this case, the Einstein equations are satisfied for,
\begin{equation}
B-C=\frac{1}{l^{2}}\; , \; A=-D=-(\frac{1}{l^{2}}+\Lambda)\label{eqA.86}
\end{equation}
where the last equality comes from equation(\ref{eqA.80}).
Then the second Yang-mills equation (\ref{eqA.52}) becomes,
\begin{equation}
B^{'}+(B+C)\lambda^{'}+\frac{1}{r}(C-D)=0\label{eqA.87}
\end{equation}
Subtracting the Bianchi-identity (\ref{eqA.43}) gives,
\begin{equation}
\frac{1}{r}(B-C+D-A)=0\label{eqA.88}
\end{equation}
So in this case, field equations are satisfied only if,
\begin{equation}
\frac{1}{r}(\frac{3}{l^{2}}+2\Lambda)=0\label{eqA.89}
\end{equation}
Thus $ r^{-1}=0\; $ implying $ B=C=0\; $ (see definitions (\ref{eqA.12}-\ref{eqA.15}), in contradiction with (\ref{eqA.84}).\\\\
\textbf{\emph{Case 3}}
\begin{equation}
B-C+\Lambda=-\frac{1}{2l^{2}}\label{eqA.90}
\end{equation}
Then (\ref{eqA.81}) is satisfied and (\ref{eqA.79}) respectively (\ref{eqA.80}) reduces to
\begin{equation}
(B+C)(3+2\Lambda l^{2})=0\label{eqA.91}
\end{equation}
\begin{equation}
(A-D)= \frac{1}{l^{2}}\label{eqA.92}
\end{equation}
So, in this case, the Einstein equations are satisfied for,
\begin{equation}
B=-C=-\frac{1}{2}(\Lambda+\frac{1}{2l^{2}})\; , \; A-D=\frac{1}{l^{2}}\label{eqA.93}
\end{equation}
Then the first Yang-mills equation (\ref{eqA.51}) becomes,
\begin{equation}
D^{'}+\frac{2}{r}(C+D +\frac{1}{l^{2}})=0\label{eqA.94}
\end{equation}
Subtracting the Bianchi-identity (\ref{eqA.44}) gives,
\begin{equation}
\frac{1}{r}(2C+\frac{1}{l^{2}})=0\label{eqA.95}
\end{equation}
Like in case 2, field equations are satisfied only if,
\begin{equation}
\frac{1}{r}(\frac{3}{l^{2}}+2\Lambda)=0\label{eqA.96}
\end{equation}
thus $r^{-1}=0$, implying $B=C=0$. Then (\ref{eqA.90}) gives $\Lambda<0$ in contradiction with observation.\\\\
This leaves case 1 as the only solution. This means that for the Schwarzschild geometry there are no solutions of the torsion free field equations if the Yang-Mills curvature energy-momentum $T^{a(\Omega)}_{\;\; b}$ is non-zero. 
 
 \newpage 
 \section{FRW Geometry}\label{App.B}
\subsection{Geometrical Quantities}
The FRW metric,
\begin{equation}
ds^{2}=-dt^{2}+a^{2}(t)\lbrace d\chi^{2}+\Sigma^{2}(\chi)(d\theta^{2}+sin^{2}\theta d\phi^{2})\rbrace
\end{equation}
with $ \Sigma(\chi)=\chi\;\;if \; k=0 ,\;\quad \Sigma(\chi)=sin\chi \;\; if \; k=1 , \; \quad  \Sigma(\chi)=sinh\chi \;\; if \; k=-1 $.\\
\\In a non-holonomic basis,
\begin{align}
ds^{2}=&\eta_{ab}\theta^{a}\wedge\theta^{b}\\
\theta^{t}=&dt\\
\theta^{\chi}=&ad\chi\\
\theta^{\theta}=&a\Sigma d\theta \\
\theta^{\phi}=&a\Sigma sin\theta d\phi
\end{align}
Calculate $ d\theta^{a}=-\omega^{a}_{\;b}\wedge\theta^{b} $ and read the connection one-forms,
\begin{align}
\omega^{\chi}\medskip_{t}=&\frac{\dot{a}}{a}\;\theta^{\chi}=\dot{a} d\chi\\
\omega^{\theta}\medskip_{t}=&\frac{\dot{a}}{a}\;\theta^{\theta}=\dot{a}\Sigma d\theta\\
\omega^{\phi}\medskip_{t}=&\frac{\dot{a}}{a}\;\theta^{\phi}=\dot{a}\Sigma sin\theta d\phi\\
\omega^{\theta}\medskip_{\chi}=&\dfrac{\Sigma^{'}}{a\Sigma}  \,\theta^{\theta} =\Sigma^{'}d\theta\\
\omega^{\phi}\medskip_{\chi}=&\dfrac{\Sigma^{'}}{a\Sigma}  \,\theta^{\phi} =\Sigma^{'} sin\theta d\phi\\
\omega^{\phi}\medskip_{\theta}=&\dfrac{cotg\theta}{a\Sigma}  \,\theta^{\phi} =cos\theta d\phi
\end{align}
Then calculate the $\Omega^{a}_{\;b}= d\omega^{a}_{\;b}+\omega^{a}_{\;c}\wedge\omega^{c}_{\;b}$ and read the Riemann-tensor components from $\Omega^{a}_{\;b}=\dfrac{1}{2}R^{a}_{\;bcd}\theta^{c}\wedge\theta^{d}$.
\begin{equation}
\Omega^{ti}=\frac{\ddot{a}}{a}\theta^{t}\wedge \theta^{i} ,\quad \Omega^{ij}=(\frac{\dot{a}^{2}}{a^{2}} +\frac{k}{a^{2}})\theta^{i}\wedge \theta^{j} \; ;\quad i,j=\chi,\: \theta,\: \phi
\end{equation}
\begin{equation}
R^{ti}\medskip_{ti}=\frac{\ddot{a}}{a}\; ,\quad R^{ij}\medskip_{ij}=\frac{\dot{a}^{2}}{a^{2}} +\frac{k}{a^{2}}\; ;\quad \,\textrm{no sum over}\;i \: \,\textrm{or}\: j,\;\; i\neq j
\end{equation}
Obtain the components of the Ricci-tensor from $ R_{ab}=R^{c}_{\;acb} $ and the Ricci-scalar $ R=R^{a}_{\;b} $
\begin{align}
R^{t}\medskip_{t}=&3\frac{\ddot{a}}{a}\: ,\quad R^{i}\medskip_{i}=\frac{\ddot{a}}{a} +2(\frac{\dot{a}^{2}}{a^{2}} +\frac{k}{a^{2}})\; , \,\textrm{no sum over}\;i\\
R=&6(\frac{\ddot{a}}{a}+\frac{\dot{a}^{2}}{a^{2}}+\frac{k}{a^{2}})
\end{align}
As for Schwarzschild geometry, we need to calculate more geometrical quantities such as the contractions of the curvature 2-form and its dual with the vector fields $ X_{a} \;$ dual to the basis 1-forms $ \theta^{a} $.\\
The contractions  $\Omega^{ab}(X_{c}) $  are obtained using $\theta^{a}(X_{b})=\delta^{a}_{\;b} $,
\begin{align}
\Omega^{ti}(X_{t})=&\frac{\ddot{a}}{a}\theta^{i}\; ,\quad \Omega^{ti}(X_{j})=-\frac{\ddot{a}}{a} \delta^{i}\medskip_{j}\theta^{t}\\
\Omega^{ij}(X_{t})=&0\; , \quad \Omega^{ij}(X_{k})=(\frac{\dot{a}^{2}}{a^{2}} +\frac{k}{a^{2}})(\delta^{i}\medskip_{k}\theta^{j}-\theta^{i}\delta^{j}\medskip_{k})
\end{align}
The dual curvature$\ast\Omega^{ab}$, is obtained using $\varepsilon_{ab}=\ast(\theta_{a}\wedge\theta_{b})=\frac{1}{2}\varepsilon_{abcd}\theta^{c}\wedge\theta^{d}$.
\begin{align}
\ast\Omega^{ti}=&-\dfrac{\ddot{a}}{a}\varepsilon_{ti}\;,\qquad 
\varepsilon_{ti}=\frac{1}{2}\varepsilon_{tijk}\theta^{j}\wedge\theta^{k}\\
\ast\Omega^{ij}=&(\frac{\dot{a}^{2}}{a^{2}} +\frac{k}{a^{2}})\varepsilon_{ij}\;,\qquad 
\varepsilon_{ij}=\varepsilon_{ijtk}\theta^{t}\wedge\theta^{k}
\end{align}
The contractions $ \ast\Omega^{ab}(X_{c}) $,  are obtained using $ \varepsilon_{ab}(X_{c})=\varepsilon_{abc}=\varepsilon_{abcd}\theta^{d} $,
\begin{align}
\ast\Omega^{ti}(X_{t})=&0 ,\quad \ast\Omega^{ti}(X_{j})=-\frac{\ddot{a}}{a} \varepsilon_{tij} , \;\varepsilon_{tij}=\varepsilon_{tijk}\theta^{k}\\
\ast\Omega^{ij}(X_{t})=&(\frac{\dot{a}^{2}}{a^{2}} +\frac{k}{a^{2}})\varepsilon_{ijt} ,\; \varepsilon_{ijt}=\varepsilon_{ijtk}\theta^{k}, \quad \ast\Omega^{ij}(X_{k})=(\frac{\dot{a}^{2}}{a^{2}} +\frac{k}{a^{2}})\varepsilon_{ijk} , \; \varepsilon_{ijk}=\varepsilon_{ijkt}\theta^{t}
\end{align}
We now have all the components to calculate the Yang-Mills energy-momentum 3-form of the gravitational field:
\begin{equation}
T_{c}\medskip^{(\Omega)}=\frac{1}{2}\lbrace \Omega^{ab}(X_{c})\wedge\ast\Omega_{ab} - \Omega^{ab}\wedge\ast\Omega_{ab}(X_{c})\rbrace
\end{equation}\\[-4mm]
Use relations as: $ \theta^{a}\wedge\varepsilon_{bc}=\delta^{a}_{\;c}\varepsilon_{b} - \delta^{a}_{\;b}\varepsilon_{c}\; $ and $\;\varepsilon_{a}=\frac{1}{3!}\varepsilon_{abcd}\theta^{b}\wedge\theta^{c}\wedge\theta^{d}\;$ to find,

\begin{align}
T_{t}\medskip^{(\Omega)}=&3\lbrace(\frac{\ddot{a}}{a})^{2}-(\frac{\dot{a}^{2}}{a^{2}}+\frac{k}{a^{2}})^{2}\rbrace\varepsilon_{t}=T_{t}\medskip^{ t (\Omega)}\varepsilon_{t}\; , \;\; \varepsilon_{t}=\frac{1}{3!}\varepsilon_{tijk}\theta^{i}\wedge\theta^{j}\wedge\theta^{k}\\
T_{i}\medskip^{(\Omega)}=&- \lbrace(\frac{\ddot{a}}{a})^{2}-(\frac{\dot{a}^{2}}{a^{2}}+\frac{k}{a^{2}})^{2}\rbrace\varepsilon_{i}=T_{i}\medskip^{ j (\Omega)}\varepsilon_{j}\; , \;\; \varepsilon_{i}-\frac{1}{2}\varepsilon_{tijk}\theta^{t}\wedge\theta^{j}\wedge\theta^{k}\\
T_{t}\medskip^{j(\Omega)}=&0 \; , \;\; T_{i}\medskip^{j(\Omega)}=0\;\; for \; \; i\neq j
\end{align}
Next we calculate the exterior covariant derivative of the dual curvature:

\begin{equation}
D\ast\Omega^{a}\medskip_{b}=d \ast\Omega^{a}\medskip_{b}+\omega^{a}\medskip_{c}\wedge \ast\Omega^{c}\medskip_{b}- \Omega^{a}\medskip_{c}\wedge \ast\omega^{c}\medskip_{b}
\end{equation}\\[-16mm]
The components are,
\begin{align}
D\ast\Omega^{t}\medskip_{i}=&(\frac{\dddot{a}}{a}+\frac{\dot{a}\ddot{a}}{a^{2}}-2\frac{\dot{a}}{a}(\frac{\dot{a}^{2}}{a^{2}}+\frac{k}{a^{2}})) \varepsilon_{i}=\frac{1}{6}\dot{R}\varepsilon_{i}\label{eqB.28}\\
D\ast\Omega^{i}\medskip_{j}=&0\label{eqB.29}
\end{align}

\subsection{Energy-Momentum and Spin Current}
Taking into account the symmetries of the FRW-geometry, the stress energy tensor complying with these symmetries, is that of a perfect fluid,
\begin{equation}
T^{(M)}_{ab}=(\rho +p)u_{a}u_{b}+p\eta_{ab}\; , \qquad u_{t}=1 , \; u_{j}=0\label{eqB.30}
\end{equation}
The fluid is completely characterised by its rest frame total (relativistic) energy density $\rho$ and isotropic pressure $p$. We assume an equation of state $p=p(\rho)$,  meaning the perfect fluid is isentropic. This energy-momentum tensor is obtained by varying the matter Lagrangian $\mathcal{L}_{M}=-\rho\varepsilon$, where $\varepsilon$ is the volume 4-form,  with respect to $ \theta^{a}$. Since $\delta_{\omega} \mathcal{L}_{M}=\frac{1}{2}\delta \omega^{ab}\wedge S_{ba}$, the spin-tensor $S_{ab}=0$.\\
A straightforward calculation of $DT^{(M)}_{a}=D_{b}T_{a}^{\;b(M)}\varepsilon$, with $D_{b}$ the covariant derivative in the non-holonomic base, shows that
\begin{equation}
DT_{t}^{(M)}=-\lbrace\dot{\rho}+3\frac{\dot{a}}{a}(\rho +p)\rbrace\varepsilon , \qquad DT_{i}^{(M)}=0\label{eqB.31}
\end{equation}
Since $ S_{ab}=0$ implies with (\ref{eq5.53}) that $ DT_{a}^{(M)}=0 $, we have the familiar conservation law
\begin{equation}
\dot{\rho}+3\frac{\dot{a}}{a}(\rho +p)=0\label{eqB.32}
\end{equation}

\subsection{Field Equations}
In the field equation (\ref{eq5.60}), we cosider as for the Schwarzschild case, the limit $l\rightarrow l_{p}$ only after solving the field equations. We also note $\Lambda=\Lambda_{obs}$, \\ 
\begin{align}
R^{a}\medskip_{b}-\frac{1}{2}R\delta^{a}\medskip_{b}+\Lambda \delta^{a}\medskip_{b}=&8 \pi l_{p}^{2}T^{a}\medskip_{b}^{(M)}-\frac{1}{2}l^{2}T^{a}\medskip_{b}^{(\Omega)}\label{eqB.33}\\
D\ast\Omega_{ab}=&0\label{eqB.34}
\end{align}
Using the curvature components and the components of the Yang-Mills energy momentum of the gravitational field, calculated in section B.1, equation (B.33) gives two Einstein equations: the $tt\:$-, and $ ii \:$-component,
\begin{align}
(\frac{\dot{a}^{2}}{a^{2}}+\frac{k}{a^{2}})-\frac{1}{2}l^{2}\lbrace(\frac{\ddot{a}}{a})^{2}-(\frac{\dot{a}^{2}}{a^{2}}+\frac{k}{a^{2}})^{2}\rbrace=&\frac{8\pi l_{p}^{2}}{3}\:\rho+\frac{\Lambda}{3}\label{eqB.35}\\
2\frac{\ddot{a}}{a}+(\frac{\dot{a}^{2}}{a^{2}}+\frac{k}{a^{2}})+\frac{1}{2}l^{2}\lbrace(\frac{\ddot{a}}{a})^{2}-(\frac{\dot{a}^{2}}{a^{2}}+\frac{k}{a^{2}})^{2}\rbrace=&-8\pi l_{p}^{2}\: p+\Lambda\label{eqB.36}
\end{align}
The $ti\:$-component of the  Yang-Mills equation (B.34), gives with (B.28), 
\begin{equation}
\dot{R}=0\label{eqB.37}
\end{equation}
Just like in Einsteins theory, the conservation equation (B.32) can be obtained from the field equations. Take the time derivative of (B.35) and then use (B.35-37). 

\subsection{Solutions}
Taking the trace of the Einstein equation leads to (or by summing (\ref{eqB.35}) and (\ref{eqB.36})),
\begin{equation}
R=8\pi l_{p}^{2}(\rho-3p)+4\Lambda\label{eqB.38}
\end{equation}
so that (\ref{eqB.37}) implies
\begin{equation}
(\rho-3p)=constant\label{eqB.39}
\end{equation}
So the Yang-Mills equation defines the equation of state. Substituting (\ref{eqB.39}) into the Einstein field equations shows that the constant term in (\ref{eqB.39}) is just an other contribution to the cosmological constant. But since $ \Lambda \; $ represents the observed cosmological constant in these field equations, we must put the constant in (\ref{eqB.39}) equal to zero. The equation of state becomes 
\begin{equation}
p=\frac{1}{3}\rho
\end{equation}
The matter content of our solutions is radiation dominated and the unique cosmological model compatible with the physical assumptions made, is the early universe.
From (\ref{eqB.32}) we now have,
\begin{equation}
\rho a^{4}=C \; ,\quad C=constant
\end{equation}
and (\ref{eqB.38}) reduces to
\begin{equation}
R=4\Lambda
\end{equation}
Solution for $ \emph{\textbf{k=0}} $.\\
On using the last two equations, the first Einstein equation reduces to,
\begin{equation}
\frac{\dot{a}^{2}}{a^{2}}-\frac{1}{3}\Lambda l^{2}(\frac{\ddot{a}}{a}-\frac{\dot{a}^{2}}{a^{2}})=\frac{8\pi l_{p}^{2}}{3}\frac{C}{a^{4}}+\frac{\Lambda}{3}
\end{equation}
which has the solution,
\begin{equation}
a^{2}=a_{i}^{2}sinh(\sqrt{\frac{4\Lambda}{3}}t)\; , \quad a_{i}^{4}=\frac{8\pi l_{p}^{2}C}{\Lambda+\frac{2}{3}l^{2}\Lambda ^{2}}
\end{equation}
where $ a_{i}=a(t_{i})\; $ , and $ \ddot{a}(t_{i})=0 $.\\
For $ a<<a_{i}\: $, we have the radiation dominated universe and the expansion is decelerated until $ a=a_{i}\: $. For $ a>a_{i}\: $, we reach the era of exponential expansion and dark energy domination. For $l=O(l_{p})\:$ (this is for a coupling constant $g_{l}$ of order unity in the Lagrangian), the solution approximates the solution from Einstein gravity.
\end{appendices}

\end{document}